\documentclass[11pt,letterpaper,twoside,reqno]{amsart}

\makeatletter{}
\usepackage[utf8]{inputenc} 
\usepackage[T1]{fontenc}    
\usepackage{url}            
\usepackage{booktabs}       
\usepackage{amsfonts}       
\usepackage{nicefrac}       
\usepackage{microtype}      
\usepackage{amsmath}

\usepackage[colorlinks=true]{hyperref}
\usepackage[dvipsnames]{xcolor}

\usepackage{graphicx}
\usepackage{epstopdf}
\usepackage{subcaption}
\usepackage{multirow}
\usepackage{float}

\usepackage{upgreek}
\usepackage{overpic}
\usepackage[]{multirow}
\usepackage[]{nicefrac}
\usepackage[]{bm}

\usepackage{tikz}
\usepackage{pgfplots}
\definecolor{skyblue1}{rgb}{0.447,0.624,0.812}
\definecolor{scarletred1}{rgb}{0.937,0.161,0.161}
\definecolor{chameleon1}{rgb}{0.541,0.886,0.204}

\newtheorem{proposition}{Proposition}
\newtheorem{definition}{Definition}

\DeclareMathOperator{\vecc}{vec}

\usepackage{xspace}
\makeatletter
\DeclareRobustCommand\onedot{\futurelet\@let@token\@onedot}
\def\@onedot{\ifx\@let@token.\else.\null\fi\xspace}

\def\ie{\emph{i.e}\onedot}

\def\etal{\emph{et al}\onedot}
\makeatother

\definecolor{astral}{RGB}{0,164,239}
\usepackage{fancyhdr}

\usepackage{titlesec}

\titleformat{\section}
{\normalfont\fontfamily{phv}\fontsize{12}{15}\bfseries\filcenter\scshape\color{astral}}{\thesection}{1em}{}

\titleformat{\subsection}
{\normalfont\fontfamily{phv}\fontsize{10}{13}\bfseries\color{astral}}{\thesubsection}{1em}{}

\titleformat{\subsubsection}
{\normalfont\fontfamily{phv}\fontsize{10}{13}\selectfont\color{astral}}{\thesubsubsection}{1em}{}

\DeclareSymbolFont{extraup}{U}{zavm}{m}{n}
\DeclareMathSymbol{\varheart}{\mathalpha}{extraup}{86}
\DeclareMathSymbol{\vardiamond}{\mathalpha}{extraup}{87}

\title[Physics-informed Autoencoders]{Physics-informed Autoencoders for Lyapunov-stable Fluid Flow Prediction}

\author[Erichson, Muehlebach, and Mahoney]{N. Benjamin Erichson$^{\LARGE \varheart}$, Michael Muehlebach$^{\LARGE \vardiamond}$ \and Michael W. Mahoney$^{\LARGE \varheart}$}

\thanks{${\Large \varheart}$ ICSI and Department of Statistics, University of California, Berkeley}
\thanks{${\Large \vardiamond}$ EECS, University of California, Berkeley}
\thanks{Corresponding author: N. Benjamin Erichson (\href{mailto:erichson@berkeley.edu}{erichson@berkeley.edu}). }

\date{}

\begin{document}

\begin{abstract}
In addition to providing high-profile successes in computer vision and natural language processing, neural networks also provide an emerging set of techniques for scientific problems.
Such data-driven models, however, typically ignore physical insights from the scientific system under consideration.
Among other things, a ``physics-informed'' model formulation should encode some degree of \emph{stability} or \emph{robustness} or \emph{well-conditioning} (in that a small change of the input will not lead to drastic changes in the output), characteristic of the underlying scientific problem. 
We investigate whether it is possible to include physics-informed prior knowledge for improving the model quality (\emph{e.g.}, generalization performance, sensitivity to parameter tuning, or robustness in the presence of noisy data). 
To that extent, we focus on the stability of an equilibrium, one of the most basic properties a dynamic system can have, via the lens of Lyapunov analysis. 
For the prototypical problem of fluid flow prediction, we show that models preserving Lyapunov stability improve the generalization error and reduce the prediction uncertainty.
\end{abstract}

\maketitle

\section{Introduction}

Many problems in science and engineering can be modeled as a dynamical system.
Examples include physical fluid flows, atmospheric-ocean interactions, neurophysiological responses, and economic and financial time series, to name only a few.
These systems often exhibit rich dynamics that give rise to multiscale structures, in both space and time.
Since these systems are typically identified using data, machine learning methods are increasingly of interest for these problems.
Deep learning and related neural network techniques, in particular, provide a useful framework for modeling such systems.
The merits of deep learning have been demonstrated for scientific applications, in particular for prototypical fluid flow applications, such as fluid flow modeling~\cite{milano2002neural,otto2019linearly,hartman2017deep,lusch2018deep,lee2018model,yeung2017learning,lui2019construction}, flow reconstruction~\cite{fukami2019super,erichson2019shallow}, flow control and prediction~\cite{morton2018deep,takeishi2017learning,vlachas2018data,puligilla2018deep,pan2018long}, and flow simulation~\cite{kim2018deep,xie2018tempogan,yeo2019deep}.

%

%

Thus far, however, neural network models for scientific applications largely ignore knowledge of physics and other domain-specific aspects of the system under consideration. 
As a consequence, this domain-agnostic learning approach can lead to models that are brittle, \emph{e.g.}, in the presence of noise, small training data, or many hyperparameters.
(This phenomenon is by now well-known for non-scientific applications such as computer vision and natural language processing.) 
One would hope that domain-specific assumptions can improve the algorithmic performance and predictive accuracy of scientific-based machine learning models.
For example, physically-informed priors can introduce some degree of \emph{stability} and \emph{robustness}, in that a small change of the input will not dramatically change the output of the learning algorithm~\cite{yu2013stability,bousquet2002stability,zheng2016improving}. 
%
%
Thus, it is no surprise that ``physics-informed'' learning has recently emerged as a topic of interest in the machine learning community~\cite{yang2018physics,wu2016physics,king2018deep,raissi2018deep,raissi2019physics,wu}, largely inspired by the pioneering work by Raissi~\etal~\cite{raissi2017physics,raissi2017physics2}. 

Motivated by this idea of physics-informed learning, we design stability-preserving models for fluid flow prediction.
This is a prototypical scientific problem with a dynamical systems interpretation.
More concretely, we learn an end-to-end mapping between the input and target fluid flow snapshot, where the mapping is represented as an autoencoder, with an additional component that attempts to learn the dynamics of the underlying physical process. 
We consider shallow network architectures (simply because shallow autoencoders show very good predictive performance for the problems of interest and are easy to train)~\cite{erichson2019shallow}.
To illustrate the promise of the method, we show results for simulated and real-world problems, including laminar flow and climate problems, and we demonstrate the use of this physics-informed approach both for improved model training and for improved \emph{a posteriori} model analysis.
For model training, we show that constraining the empirical risk minimization problem by using a Lyapunov stability-promoting prior (a physical-meaningful regularization mechanism that corresponds to properties of the physical system being modeled) leads to better training and helps to improve the generalization performance, compared to physics-agnostic models. 
In other words, we introduce a physical-meaningful regularization mechanism to ensure a better generalization performance. 
The extra tuning parameter for promoting stability can be regarded as a parameter that controls the fitting-stability trade-off. 
%
%


\section{Problem setup}

We assume that the dynamic system of interest can be modeled as
\begin{equation}\label{eq:dynamical_model}
\mathbf{x}_{t+1} =  \mathcal{A}(\mathbf{x}_t)+\eta_t, \quad t=0,1,2, \dots,
\end{equation}
where $\mathbf{x}_t$ denotes the state of the system at time $t$, where $t$ is a non-negative integer. The function $\mathcal{A}:\mathbb{R}^n \rightarrow \mathbb{R}^n$ maps the state at time $t$ to the state at time $t+1$, and $\eta_t$ is a (small) perturbation. The perturbation $\eta_t$ might incorporate modeling errors, such as slowly changing operating conditions (unmodeled dynamics) or discretization errors. 
If $\eta_t$ is small, then the dynamics simply specify here that the state $\mathbf{x}_{t+1}$ depends just on the value of the previous state $\mathbf{x}_t$, i.e., given the rule $\mathcal{A}$, the state $\mathbf{x}_{t}$ provides all information needed for predicting the future state at $\mathbf{x}_{t+1}$.

Modeling nonlinear dynamics can be challenging and thus often linear time-invariant approximations of nonlinear systems are used.
These take the form
\begin{equation}\label{eq:dynamical_model_linear}
\mathbf{x}_{t+1} =  \mathbf{A}\mathbf{x}_t + \eta_t, \quad t=0,1,2, \dots,
\end{equation}
where $\mathbf{A}:\mathbb{R}^n \rightarrow \mathbb{R}^n$ denotes a linear map.
Linear models provide a reasonably good approximation for the underlying process in many applications. 
Thus, for simplicity, we assume that the underlying dynamics of the problems which we consider in the following are linear.
However, despite the simplicity of this rule, it often turns out to be a challenge to find an estimate for $\mathbf{A}$. This is because, in a data-driven setting, we only have access to (high-dimensional) observations
\begin{equation}\label{eq:output}
\mathbf{y}_t = \mathcal{G}(\mathbf{x}_t)+\mathbf{\xi}_t, \quad t=0,1,2,\dots,T,
\end{equation}
where the function $\mathcal{G}: \mathbb{R}^n \rightarrow \mathcal{Y} \subseteq \mathbb{R}^m$ maps the state $\mathbf{x}_t$ to a subspace $\mathcal{Y}$, and the variable $\mathbf{\xi}_t$ represents measurement errors. For example, one may think of the function $\mathcal{G}$ as a sensor which collects measurements at time $t$. It is assumed that the integer $m$ is much larger than $n$, \ie, we assume that the dynamics of the flow are low-dimensional, in the sense that $\mathcal{G}(\mathbf{x}_t)$, $t=1,2,\dots$, lies on an n-dimensional manifold embedded in $\mathbb{R}^m$~\cite{benner2015survey}. 
Further, we assume that the function $\mathcal{G}$ has an inverse, which implies that a single data-point $y_t\in \mathcal{Y}$ is enough to uniquely determine the corresponding state $\mathbf{x}_t$.
Provided that the dynamics are observable, this assumption can always be met by vertically stacking several snapshots together~\cite{sastry2013nonlinear}. 

For our fluid flow example (in Sec.~\ref{sec:experiments}), the state $\mathbf{x}_t$ might represent a spatial discretization of the velocity and pressure field, and the output $\mathbf{y}_t$ a spatial discretization of the velocity flow field. 
The variables $\eta_t$ capture, \emph{e.g.}, unknown temperature variations, change of boundary conditions, and spatial discretization errors. The variables $\xi_t$ model spatial discretization and quantization errors. 

%
%
%

\section{Autoencoder-type models for fluid flow prediction}

Given a sequence of observations $\mathbf{y}_0,\mathbf{y}_1, \dots, \mathbf{y}_T \in \mathbb{R}^m$ for training, the objective of this work is to learn a model which maps the snapshot $\mathbf{y}_t$ to $\mathbf{y}_{t+1}$. The model is composed of three functions
\begin{equation}\label{eq:flowmap_model}
\mathbf{\hat{y}}_{t+1} =  \mathbf{\Phi} \circ \mathbf{\Omega} \circ \mathbf{\Psi}(\mathbf{y}_t),
\end{equation}
where $\mathbf{\Phi}$ approximates $\mathcal{G}$, $\mathbf{\Omega}$ approximates $\mathbf{A}$, and $\mathbf{\Psi}$ approximates $\mathcal{G}^{-1}$.
%

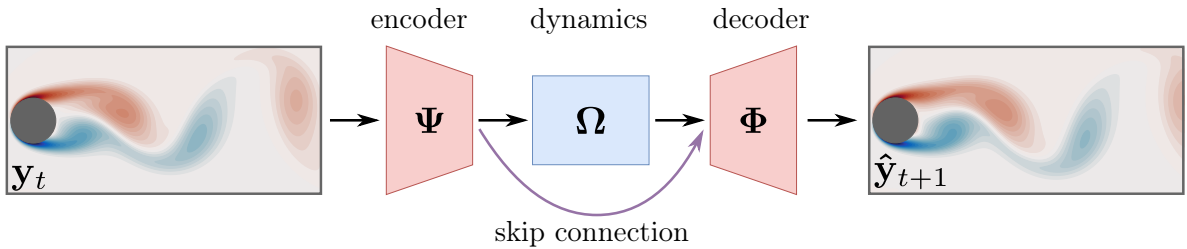
\begin{figure}[!b]
	\centering
	\DeclareGraphicsExtensions{.pdf}
	\vspace{+0.5cm}
	\begin{overpic}[width=0.95\textwidth]{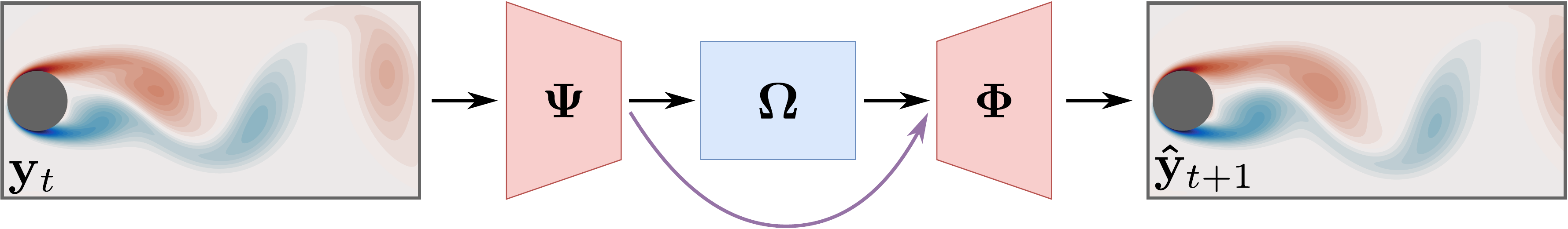}
		\put(31,16){{encoder}}
		\put(44.5,16){{dynamics}}	
		\put(60,16){{decoder}}
		\put(41.5,-2){{skip connection}}				
	\end{overpic}\vspace{+0.4cm}	
	\caption{Design architecture of the autoencoder-type flow prediction model. The skip connection allows one to enforce the identity-persevering constraint posed on the encoder. This constraint is important, because we aim to design the model so that only $\mathbf{\Omega}$ captures the dynamics. }
	\label{fig:overview}
\end{figure}

The encoder $\mathbf{\Psi}: \mathcal{Y} \rightarrow \mathbb{R}^n$ maps the high-dimensional snapshot $\mathbf{y}_t$ to a low-dimensional feature space, where $n\ll m$. The encoder should be designed so that it preserves the coherent structure of the fluid flow, while suppressing uninformative variance (fine scale features) in the data.
The dynamics $\mathbf{\Omega}: \mathbb{R}^n \rightarrow \mathbb{R}^n$ evolves the state in time, modeled as 
\begin{equation}\label{eq:rom_model_linear}
\mathbf{z}_{t+1} =  \mathbf{\Omega}\mathbf{z}_t.
\end{equation}
where $\mathbf{z}_t = \mathbf{\Psi}(\mathbf{y}_t)$.
Note that this layer poses a bottleneck, \emph{i.e.}, the size $k$ of the feature space constraints the expressive capacity of the model. 
Finally, the decoder $\mathbf{\Phi}: \mathbb{R}^n \rightarrow \mathcal{Y}$ maps the low-dimensional features (evolved in time) back to the high-dimensional measurement space. 
The design architecture of the model is sketched in Figure~\ref{fig:overview}.

Given the pairs $\{\mathbf{y}_{t},\mathbf{y}_{t+1} \}_{t=1,2,...}$, we train the model by minimizing the mean squared error
\begin{equation}\label{eq:base_loss}
\min \frac{1}{T-1}\sum_{t=0}^{T-1} \,\, \|\mathbf{y}_{t+1} - \mathbf{\Phi} \circ \mathbf{\Omega} \circ \mathbf{\Psi}(\mathbf{y}_t) \|_2^2.
\end{equation}
%

During inference time we can obtain predictions $\hat{\mathbf{y}}_t$ for an initial point $\mathbf{y}_0$ by composing the learned model $t$-times, \ie, given $\mathbf{y}_0$ we use the output $\mathbf{y}_{1}$ as input to predict $\mathbf{y}_{2}$, and so on. This leads to the following expansion
\begin{align}\label{eq:auto_iterate}
\mathbf{\hat{y}}_{t} = \mathbf{\Phi} \circ \mathbf{\Omega} \circ \mathbf{\Psi}  \circ \mathbf{\Phi} \circ \mathbf{\Omega} \circ \mathbf{\Psi}  \circ \mathbf{\Phi} \circ \mathbf{\Omega}  \circ \mathbf{\Psi}  \circ \,...\,  \circ \mathbf{\Phi} \circ \mathbf{\Omega} \circ \mathbf{\Psi}(\mathbf{y}_0).
\end{align}

If the model obeys the assumption that $\mathbf{\Psi}$ approximates $\mathcal{G}^{-1}$, then we have that $\mathbf{I} \approx \mathbf{\Psi}  \circ \mathbf{\Phi}$, where $\mathbf{I} \in \mathbb{R}^{n\times n}$ denotes the identity. Thus, Eq.~\eqref{eq:auto_iterate} reduces approximately to
\begin{align}\label{eq:predictions}
\mathbf{\hat{y}}_{t} \approx \mathbf{\Phi} \circ \mathbf{\Omega} \circ \mathbf{\Omega} \circ \mathbf{\Omega} \,...\, \circ \mathbf{\Omega} \circ \mathbf{\Psi}(\mathbf{y}_0) = \mathbf{\Phi} \circ \mathbf{\Omega}^t \circ \mathbf{\Psi}(\mathbf{y}_0).
\end{align}

In order for this to happen, we enforce that the function $\mathbf{\Psi}$ acts as an (approximate) inverse function for $\mathbf{\Phi}$, i.e., $\mathbf{q}_t \approx \mathbf{\Psi}  \circ \mathbf{\Phi}(\mathbf{q}_t)$. This is achieved by introducing an additional penalty 
\begin{equation}\label{eq:improve_loss}
\min \frac{1}{T-1}\sum_{t=0}^{T-1} \,\, \|\mathbf{y}_{t+1} -  \mathbf{\Phi} \circ \mathbf{\Omega} \circ \mathbf{\Psi}(\mathbf{y}_t) \|_2^2 \,\, + \,\, \lambda \| \mathbf{q}_t - \mathbf{\Psi} \circ \mathbf{\Phi}(\mathbf{q}_t)  \|_2^2,
\end{equation}
where $\lambda$ is a tuning parameter that balances the two objectives. The variable $\mathbf{q}_t$ denotes carefully chosen test points, which we set in our experiments to the encoded flow field $\mathbf{q}_t = \mathbf{\Psi}(\mathbf{y}_t)$ at time $t$.

%

\section{Lyapunov stability as a tool for physics-informed learning}

Depending on the context, different notions of stability may be appropriate. 
For instance, the stability of a statistical procedure is typically understood as robustness with respect to small perturbations of the data~\cite{yu2013stability,jen2003stable}. Stability in machine learning can mean that small perturbations of certain tuning parameters, such as amount of weight decay, ratio of dropout and learning rate, lead to models that yield similar generalization errors~\cite{bousquet2002stability,Hardt}. 

Here, we focus on Lyapunov stability, which describes a fundamental property of the dynamic system in Eq.~\eqref{eq:dynamical_model}. We assume that the dynamic system has an equilibrium at the origin, \emph{i.e.}, $\mathcal{A}(0)=0$.
This means that if the system is initialized at the origin ($\mathbf{x}_0=0$), its state will remain at the origin for all times (that is, $\mathbf{x}_t=0$ for all $t=1,2,\dots$ provided that $\eta_t=0$ for all $t=1,2,\dots$).
We can then ask ourselves what happens when the system is initialized in a region close to the the origin. Will the resulting trajectories remain close to the equilibrium for all times, or will they drift away? If the former is true, the origin of the dynamic system is said to be stable. Compared to the other notions of stability mentioned above, Lyapunov stability is therefore a statement about the robustness of trajectories with respect to small perturbations of their initial conditions about the given equilibrium. 
Note that we always refer to an equilibrium when introducing the notion of Lyapunov stability. In case the dynamics are linear such as in Eq.~\eqref{eq:rom_model_linear}, we assume (often without saying) that the equilibrium in question is the origin.


Provided that the origin of the system~\eqref{eq:dynamical_model} is stable, we would like to ensure that the origin of the approximation $\mathbf{\Omega}$ in Eq.~\eqref{eq:rom_model_linear} is likewise stable. 
The two main reasons for this are the following.

\begin{itemize}
	\item[1)] If the resulting approximation $\mathbf{\Omega}$ is not stable, even tiny perturbations of $\hat{\mathbf{y}}_0$ might lead to unbounded output predictions $\hat{\mathbf{y}}_k$ (according to Eq.~\eqref{eq:predictions}), which might thus be arbitrarily far off. However, if the approximation $\mathbf{\Omega}$ is stable, then we at least know that small perturbations of $\hat{\mathbf{y}}_0$ about the origin lead to bounded output predictions $\hat{\mathbf{y}}_k$.  
	
	\item[2)] Imposing that the underlying dynamics are stable can be viewed as a basic, but principled regularization. We show with numerical examples that this improves the generalization performance and reduces the sensitivity with respect to hyperparameters.
\end{itemize}

The following sections provide some background information on Lypaunov stability and introduce several approaches for imposing the additional requirement of $\mathbf{\Omega}$ being stable.

\subsection{Lyapunov stability}

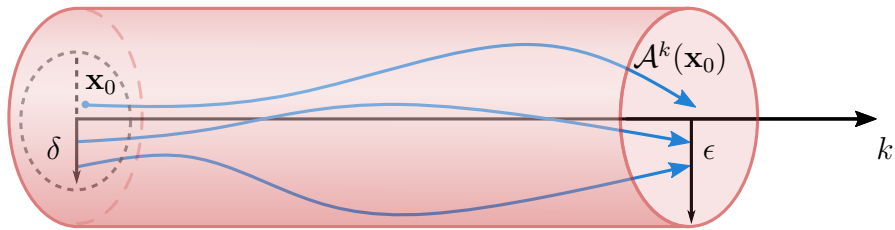
\begin{figure}[!b]
	\centering
	\DeclareGraphicsExtensions{.pdf}
	\begin{overpic}[width=0.7\textwidth]{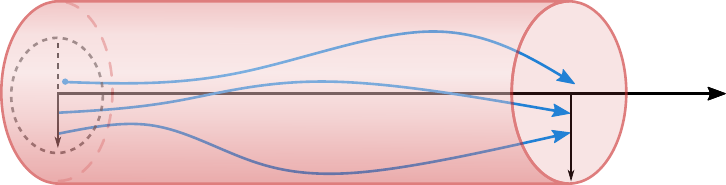} 
		\put(4.5,8){\large {$\delta$}}
		\put(80,8){\large {$\epsilon$}}
		\put(100,8){\large {$k$}}
		\put(9,16){\large {$\mathbf{x}_0$}}
		\put(72,18.5){\large {$\mathcal{A}^k (\mathbf{x}_0)$}}
	\end{overpic}
	\caption{Illustration of the idea of Lyapunov stability.}
	\label{fig:illustration_lyap}
\end{figure}

There are many textbooks that describe Lyapunov stability~\cite{hahn1967stability, sastry2013nonlinear}. For the sake of completeness, here we provide a short summary of the results that will be used subsequently.
\begin{definition}\label{Def:Stab}
	The origin is a stable equilibrium of \eqref{eq:dynamical_model} if for every $\epsilon>0$ there exists a $\delta > 0$ such that $|\mathbf{x}_0|<\delta$ implies
	\begin{equation}\label{eq:stabdef}
	|\mathcal{\mathcal{A}}^k (\mathbf{x}_0)|<\epsilon, \quad \text{for~all~}k=1,2,3,\dots.
	\end{equation}
\end{definition}
In other words, the origin is stable if all trajectories starting arbitrarily close to the origin (in a ball of radius $\delta$) remain arbitrarily close (in a ball of radius $\epsilon$).
This is illustrated in Figure~\ref{fig:illustration_lyap}. 
We say that the origin is \emph{asymptotically stable} if, in addition to \eqref{eq:stabdef}, it holds that
\begin{equation}\label{eq:conv}
\lim_{k\rightarrow\infty} \mathcal{A}^k(\mathbf{x}_0) =0.
\end{equation}
This means that the trajectories starting close to the equilibrium not only remain close, but eventually converge to the origin. We should emphasize, however, that the requirement \eqref{eq:conv} alone is not enough for ensuring asymptotic stability, as there are examples where trajectories starting arbitrary close to the equilibrium meander arbitrarily far before converging. In such a case, the origin is not asymptotically stable, even though \eqref{eq:conv} is fulfilled for any initial condition $\mathbf{x}_0$ close to the origin.

In general, it may be very difficult to determine whether the equilibrium of a given system is stable, as this includes the computation of the trajectories $\mathcal{A}^k(\mathbf{x}_0)$ for all future times and for all initial conditions $\mathbf{x}_0$ in a region about the origin. This is where the so-called second method of Lyapunov provides us with an elegant solution. 
It can be shown that the origin is \emph{stable in the sense of Lyapunov} if there exists a continuous function $\mathcal{V}:\mathbb{R}^n\rightarrow \mathbb{R}$ such that
\begin{align}
&\mathcal{V}(0)=0, \quad \mathcal{V}(\mathbf{x})>0, \quad \text{for~all~$\mathbf{x}\neq 0$}, \label{eq:Lyap1}\\ &\mathcal{V}(\mathcal{A}(\mathbf{x}))-\mathcal{V}(\mathbf{x})\leq 0, \quad \text{for~all~$\mathbf{x}\neq 0$, $|\mathbf{x}|<r$}, \label{eq:Lyap2}
\end{align}
where $r$ is a (possibly small) positive constant. 
The origin is \emph{asymptotically stable in the sense of Lyapunov} if the inequality in \eqref{eq:Lyap2} is strict. 
The function $\mathcal{V}$ can viewed as generalized ``energy function,'' since it is often related to the total energy of the dynamical system \eqref{eq:dynamical_model}. Thus the statement implies that the origin is stable if there exists an ``energy function'' that decreases along all trajectories that are close to the origin (\emph{i.e.}, the system dissipates energy). In order to check \eqref{eq:Lyap1} and \eqref{eq:Lyap2}, we are not required to compute the full trajectories $\mathcal{A}^k(\mathbf{x}_0)$.
This is why the second method of Lyapunov is a powerful tool and frequently used for the analysis and control of dynamical systems.

In case the dynamics are linear, (Lyapunov) stability of the origin can be checked with an eigenvalue analysis. More precisely, the following two results can be deduced from a Jordan decomposition of the matrix $\mathbf{A}$ combined with the Definition~\ref{Def:Stab}.
\begin{itemize}
	\item The origin of the dynamical system~\eqref{eq:dynamical_model_linear} is stable in the sense of Lyapunov if and only if all eigenvalues of $\mathbf{A}$ have magnitude less than one and the eigenvalues with magnitude one have equal algebraic and geometric multiplicity. 
	\item The origin of the dynamical system~\eqref{eq:dynamical_model_linear} is asymptotically stable if and only if all eigenvalues of $\mathbf{A}$ have magnitude strictly less than one.
\end{itemize}
In both cases, the fact that the linear dynamics are stable (or asymptotically stable), implies that all trajectories remain bounded. Thus, provided that $\mathbf{\Psi}$ and $\mathbf{\Phi}$ are bounded maps, we are guaranteed to obtain bounded predictions $\hat{\mathbf{y}}_t$ for all times $t=0,1,\dots$.
For linear systems, Lyapunov's second method reduces to the following statement,~\cite{agarwal2000difference}(p.~285, Thm.~5.9.4)
\begin{proposition}\label{Prop:Lyap}
	The origin of the dynamic system~\eqref{eq:dynamical_model_linear} is asymptotically stable in the sense of Lyapunov if and only if for any (symmetric) positive definite matrix $\mathbf{Q}\in \mathbb{R}^{n\times n}$ there exists a (symmetric) positive definite matrix $\mathbf{P} \in \mathbb{R}^{n\times n}$ satisfying
	\begin{equation}\label{eq:Lyap}
	\mathbf{A}^{\top} \mathbf{P} \mathbf{A} - \mathbf{P} = -\mathbf{Q}.
	\end{equation}
\end{proposition}
The proposition can be proved by noting the following.
\begin{itemize}
	\item Provided that $\mathbf{P}$ satisfies \eqref{eq:Lyap} and is positive definite, $\mathcal{V}(\mathbf{x})=\mathbf{x}^{\top} \mathbf{P} \mathbf{x}$ is a Lyapunov function satisfying \eqref{eq:Lyap1} and \eqref{eq:Lyap2} (in a strict sense).
	
	\item Provided that the dynamic system~\eqref{eq:dynamical_model_linear} is asymptotically stable, $\mathbf{P}=\sum_{k=0}^{\infty} (\mathbf{A}^{\top})^k \mathbf{Q} \mathbf{A}^k$ is positive definite and satisfies \eqref{eq:Lyap}.
\end{itemize}
The formulation via Prop.~\ref{Prop:Lyap} is useful in many applications, since checking the stability of $\mathbf{A}$ is reduced to checking whether the matrix $\mathbf{P}$ defined by \eqref{eq:Lyap} is positive definite. The matrix $\mathbf{P}$ is symmetric, and it has therefore real eigenvalues, which simplifies the computation of gradients. Without loss of generality, we define $\mathbf{Q}$ to be the identity $\mathbf{I}$, which satisfies the condition that $\mathbf{Q}$ is positive definite.

\subsection{Physics-informed models based on Lyapunov's method}

Given that~\eqref{eq:dynamical_model_linear} is stable, we are interested in learning a model~\eqref{eq:rom_model_linear} that is likewise stable. 
Therefore, we design a stability-promoting penalty based on Lyapunov's method as formulated by Prop.~\ref{Prop:Lyap}.
More precisely, we impose that the symmetric matrix $\mathbf{P}$, defined by
\begin{equation}\label{eq:LyapI}
\mathbf{\Omega}^{\top} \mathbf{P} \mathbf{\Omega} - \mathbf{P} = -\mathbf{I},
\end{equation}
is positive definite. 

To gain some intuition for the functional relationship of the eigenvalues between $\mathbf{\Omega}$ and $\mathbf{P}$, we consider the case where $\mathbf{\Omega}$ is diagonalizable and $\mathbf{Q}$ chosen appropriately.
Then, for a particular choice of coordinates,~\eqref{eq:LyapI} reduces to the system of linear equations $\omega_i p_i \omega_i - p_i = -1$, where $\omega_i$, $p_i$, for $i=1,2,\dots,n$, denote the eigenvalues of $\mathbf{\Omega}$ and $\mathbf{P}$, respectively.
Then, we can solve for $\omega_i$ which yields
\begin{equation}
\omega_i = \pm \sqrt{1-\frac{1}{p_i}}.
\end{equation}
Figure~\ref{fig:lyap1d} shows the functional relationship between $\omega_i$ and $p_i$.
Prop.~\ref{Prop:Lyap} states that $\omega_i$ has magnitude less than one if and only if $p_i$ is greater than one, as can be verified in Figure~\ref{fig:lyap1d}.
Moreover, $p_i$ takes values close to zero when $\omega_i$ takes large values (rapidly diverging dynamics), whereas $p_i$ becomes very small ($\rightarrow -\infty$), when $\omega_i$ takes values with magnitude slightly larger than one.

Motivated by this observation, we design a prior that promotes Lyapunov stability by penalizing eigenvalues $p$ that have small negative values. Such a prior can take various forms, but the following choice (illustrated in Fig.~\ref{fig:prior}) works particularly well in our experiments:
\begin{equation}
\label{eqn:rho_function}
\rho(p) := \begin{cases}
\exp\left(-\frac{|p-1|}{\gamma}\right) \quad \text{if} \,\,\, p<0 \\
0 \quad \quad \quad \quad \quad \,\,\,\, \text{otherwise},
\end{cases}
\end{equation}
where $\gamma$ is a tuning parameter (we set to $\gamma=4$ in our experiments). The physics-informed autoencoder is then trained by minimizing the following objective 
\begin{equation}\label{eq:pcl_loss}
\min  \frac{1}{T-1}\sum_{t=0}^{T-1} \,\, \|\mathbf{y}_{t+1} -  \mathbf{\Phi} \circ \mathbf{\Omega} \circ \mathbf{\Psi}(\mathbf{y}_t) \|_2^2 \,\, + \,\, \lambda \| \mathbf{q}_t - \mathbf{\Psi} \circ \mathbf{\Phi}(\mathbf{q}_t)  \|_2^2 + \,\,  \kappa \sum_i \rho(p_i) ,
\end{equation}
where $\rho(\cdot)$ is given by Eq.~(\ref{eqn:rho_function}).
This preserves stability if $\kappa$ is chosen large enough.

\pgfplotsset{
	double y domain/.code 2 args={
		\pgfmathsetmacro\doubleymin{#1*2}
		\pgfmathsetmacro\doubleymax{#2*2}
	},
	ejes/.style args={#1:#2 #3:#4}{
		double y domain={#3}{#4},
		domain=#1:#2,
		ymin=#3,ymax=#4, restrict y to domain=\doubleymin:\doubleymax,
		samples=100,
		enlargelimits=false,
		axis lines=middle,
		xtick={#1,...,#2}, ytick={#3,...,#4},
		xticklabels=\empty, yticklabels=\empty,
		every axis plot post/.style={
			mark=none,
			smooth
		},
		scale only axis,
		width=6cm,
		height=4cm
	}
}

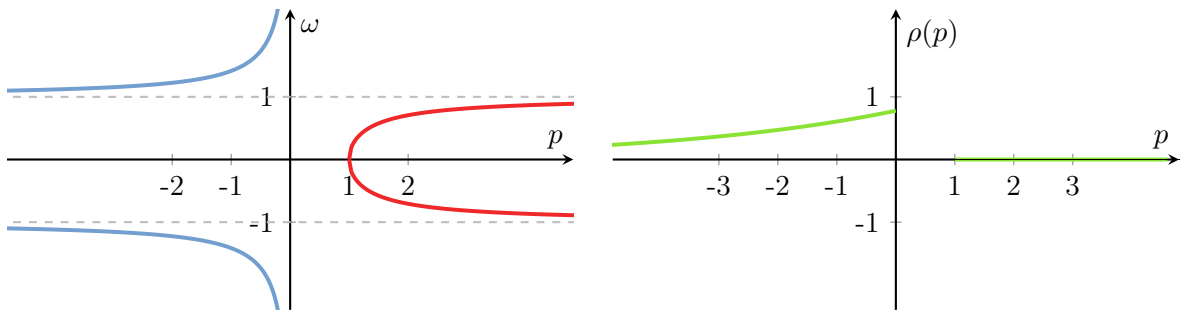
\begin{figure}[!b]
	\centering
	
	\begin{subfigure}[t]{0.48\textwidth}
		\centering
		\tikz{\begin{axis}[ejes=-9:9 -2.:2., xlabel={$p$}, ylabel={$\omega$},
				thick,
				axis lines = middle,
				enlargelimits = true,
				width=.95\textwidth, height=4cm,
				xmin=-4, xmax=4, ymin=-2., ymax=2.,
				no markers,
				samples=200,
				axis lines*=left, 
				axis lines*=middle, 
				scale only axis,
				xtick={-2, -1,  0,  1, 2},
				ytick={-1, 0, 1},
				xticklabels={-2, -1, 0, 1, 2},
				yticklabels={-1, 0, 1}
				]
				
				\addplot[line width=1.5pt, color=scarletred1, domain=1:8] {sqrt((1-1/x))};
				\addplot[line width=1.5pt, color=scarletred1, domain=1:8] {-sqrt((1-1/x))};
				
				\addplot[line width=1.5pt, color=skyblue1, domain=-8:0] {sqrt((1-1/x))};
				\addplot[line width=1.5pt, color=skyblue1, domain=-8:0] {-sqrt((1-1/x))};		
				
				\addplot[mark=none, lightgray, dashed] {1};
				\addplot[mark=none, lightgray, dashed] {-1};	

		\end{axis}}
		\caption{Discrete-time Lyapunov function.}\label{fig:lyap1d}
	\end{subfigure}
	~
	\begin{subfigure}[t]{0.48\textwidth}
		\centering
		\tikz{\begin{axis}[ejes=-9:9 -2.:2., xlabel={$p$}, ylabel={$\rho(p)$},
				thick,
				axis lines = middle,
				enlargelimits = true,
				width=.95\textwidth, height=4cm,
				xmin=-4, xmax=4, ymin=-2., ymax=2.,
				no markers,
				samples=200,
				axis lines*=left, 
				axis lines*=middle, 
				scale only axis,
				xtick={-3, -2,  -1,  0, 1,  2, 3},
				ytick={-1, 0, 1},
				xticklabels={-3, -2, -1,  0, 1, 2, 3},
				yticklabels={-1, 0, 1}
				]
				
				\addplot[line width=1.5pt, color=chameleon1, domain=-8:0] {exp(- abs(x-1) / 4)};
				\addplot[line width=1.5pt, color=chameleon1, domain=1:4.6] {0};
				\addplot[line width=0.0pt, color=black, domain=1:8] {0};
				
				
		\end{axis}}
		\caption{Stability promoting prior.}\label{fig:prior}
	\end{subfigure}
	
	\caption{The relationship between the eigenvalues of $\mathbf{\Omega}$ and $\mathbf{P}$ is shown in (a). In (b) our choice for a stability promoting prior is shown. The prior emphasizes on penalizing smaller eigenvalues of $\mathbf{P}$ less than larger eigenvalues as long as $p_i \leq 0$.}
	\label{fig:lyap}
\end{figure}

The challenge remains to compute the eigenvalues of $\mathbf{P}$ given the matrix $\mathbf{\Omega}$.
%
We can use the Kronecker product to transform Eq.~(\ref{eq:Lyap}) into the equivalent system of linear equations taking the form
\begin{equation}\label{eq:minkronecker}
(\mathbf{\Omega}^{\top} \otimes \mathbf{\Omega}^{\top}) \,  \vecc{(\mathbf{P})}  - \vecc{(\mathbf{P})} = \vecc{(-\mathbf{I})},
\end{equation}
where $\otimes$ denotes the Kronecker product~\cite{van2000ubiquitous}, and the $\vecc$-operator stacks the entries of a $n\times n$ matrix columnwise into a vector of length $n^2$. We can determine $\mathbf{P}$ by solving the system of linear equations (which might have multiple solutions) using the Moore-Penrose inverse~\cite{penrose1955generalized}. Specifically, all solution for $\vecc{(\mathbf{P})}$ are given by
\begin{equation}
\vecc{(\mathbf{P})}  = (\mathbf{\Omega}^{\top} \otimes \mathbf{\Omega}^{\top} - \mathbf{I})^+  \vecc{(-\mathbf{I})} + \left[\mathbf{I} -  
(\mathbf{\Omega}^{\top} \otimes \mathbf{\Omega}^{\top} - \mathbf{I})^+ (\mathbf{\Omega}^{\top} \otimes \mathbf{\Omega}^{\top} - \mathbf{I})  \right]\mathbf{v},
\end{equation}  
where $^+$ denotes the pseudoinverse, and $\mathbf{v}$ denotes an arbitrary vector.
Here we make the assumption that $(\mathbf{\Omega}^{\top} \otimes \mathbf{\Omega}^{\top} - \mathbf{I})$ is a full rank square matrix, so that we can obtain a unique solution by computing
\begin{equation}\label{eq:solution}
\vecc{(\mathbf{P})}  = (\mathbf{\Omega}^{\top} \otimes \mathbf{\Omega}^{\top} - \mathbf{I})^+  \vecc{(-\mathbf{I})}.
\end{equation} 
In practice, however, it is costly to form the Kronecker product in Eq.~\eqref{eq:solution} explicitly. 
Instead, one can compute the QR decomposition of $\mathbf{\Omega}$ and then exploit the properties of the Kronecker product to compute a solution for Eq.~\eqref{eq:minkronecker}; see~\cite{fausett1994large,hammarling1982numerical}. An alternative robust implementation to solve the Lyapounov discrete equation is based on the Schur decomposition, proposed by Barraud~\cite{4045877}. 


The derivative of the eigenvalues $\lambda_i$ and eigenvectors $\mathbf{v}_i$ of $\mathbf{P}$  can now be computed by~\cite{magnus1985differentiating}, as
\begin{equation}
\partial \lambda_i = \mathbf{v}_i^\top \partial (\mathbf{P}) \mathbf{v}_i,
\end{equation} 
and
\begin{equation}
\partial \mathbf{v}_i = (\lambda_i \mathcal{I} - \mathbf{P})^+ \partial (\mathbf{P}) \mathbf{v}_i,
\end{equation} 
provided that all eigenvalues are distinct. These formulas exploit the fact that $\mathbf{P}$ is real and symmetric.

\section{Experiments and discussion}\label{sec:experiments}

In the previous section, we introduced a methodology to design physics-informed autoencoders that preserve stability. 
In this section, we provide empirical results, demonstrating the generalization performance, by studying a periodic flow behind a cylinder and real-world climate problem.

We use shallow architectures which are composed of only a few linear layers, connected by non-linear activation functions.   These architectures provide an excellent parsimonious-predictability trade-off for our fluid flow prediction problems. In addition, shallow networks have the advantage that they are scalable, fast to train, and easy to tune~\cite{erichson2019shallow}. 
We use the tanh-activation function, because it shows a better performance than the ReLU for our problems.
Additional results and specifics of the network architecture are presented in the supplementary materials.

\begin{figure}[!b]
	\centering
	\begin{subfigure}[t]{0.3\textwidth}
		\centering
		\DeclareGraphicsExtensions{.pdf}
		\includegraphics[width=1\textwidth]{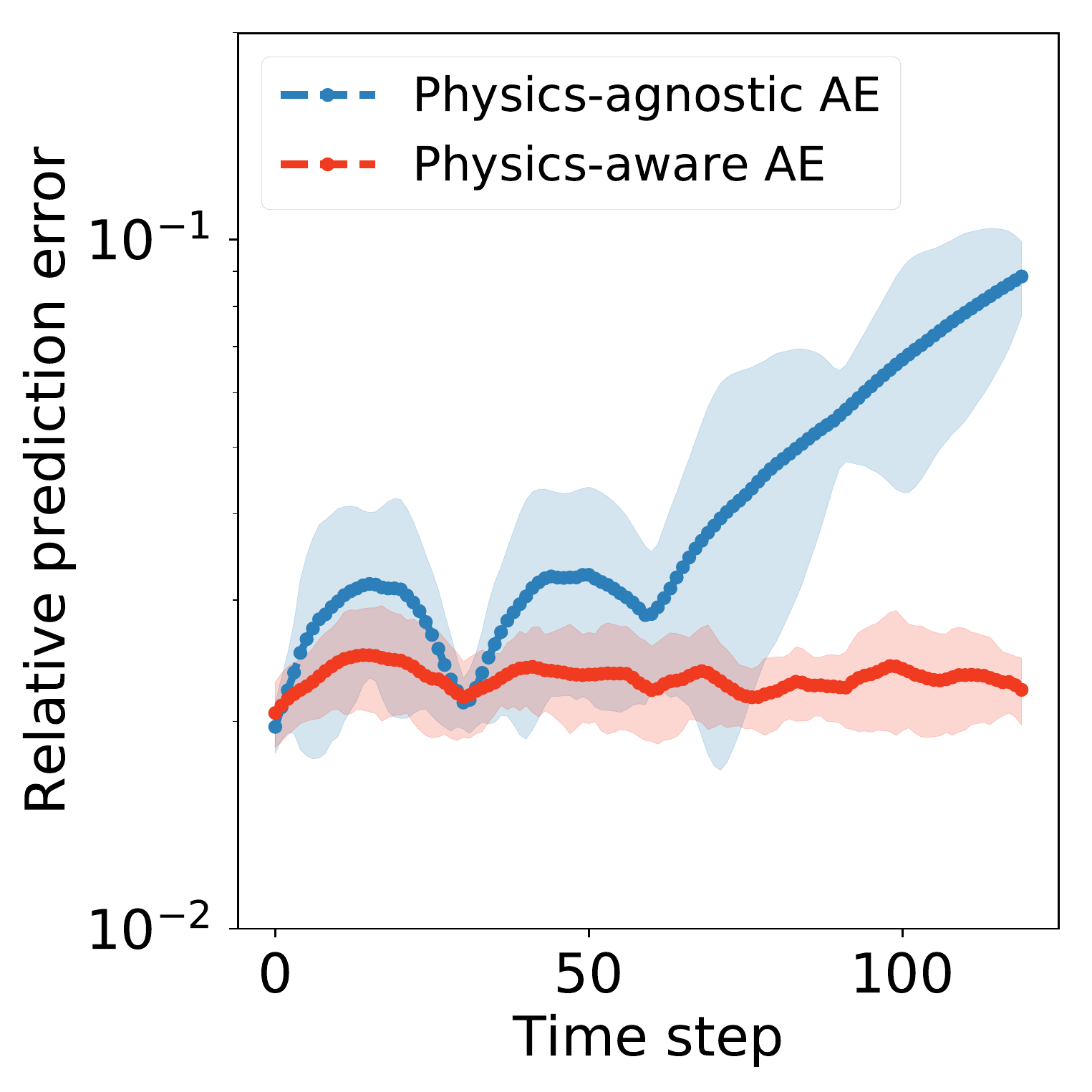}
		\caption{With LR $1\mathrm{e}{-2}$ and WD $1\mathrm{e}{-6}$.}
	\end{subfigure}
	~
	\begin{subfigure}[t]{0.3\textwidth}
		\centering
		\DeclareGraphicsExtensions{.pdf}
		\includegraphics[width=1\textwidth]{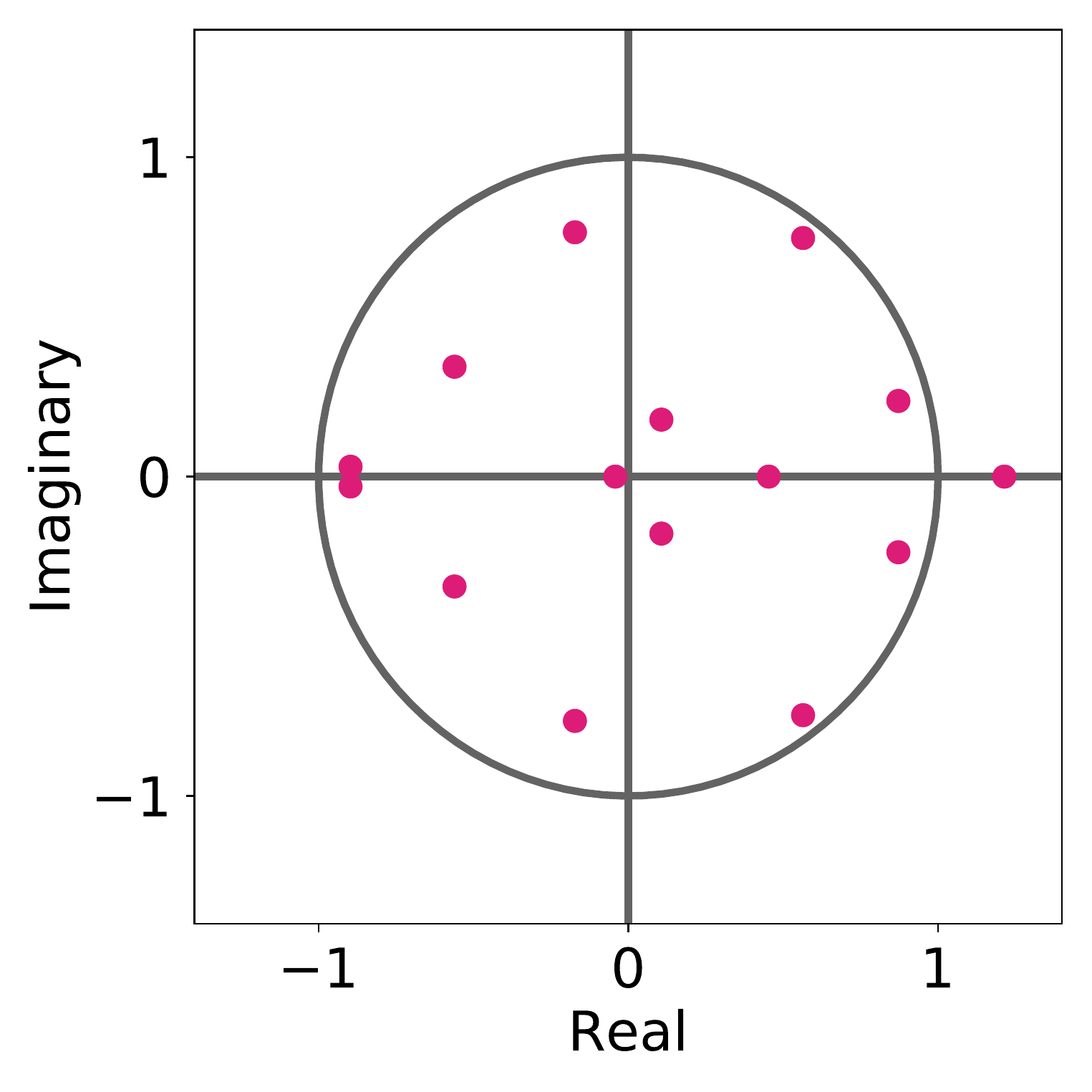}
		\caption{Physics-agnostic model in (a).}
	\end{subfigure}\vspace{+0.1cm}
	~
	\begin{subfigure}[t]{0.3\textwidth}
		\centering
		\DeclareGraphicsExtensions{.pdf}
		\includegraphics[width=1\textwidth]{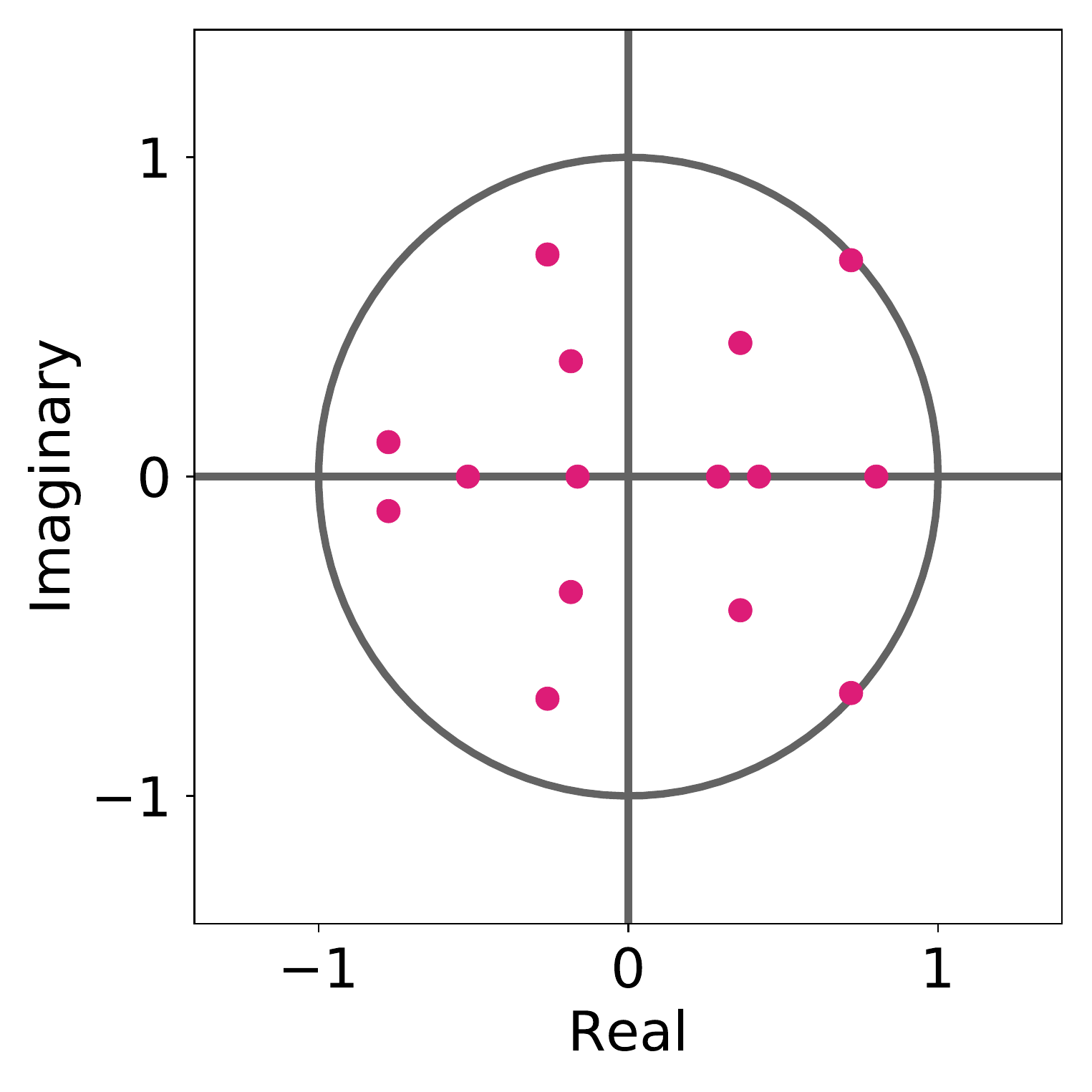}
		\caption{Physics-aware model in (a).}
	\end{subfigure}
	
	\begin{subfigure}[t]{0.31\textwidth}
		\centering
		\DeclareGraphicsExtensions{.pdf}
		\includegraphics[width=1\textwidth]{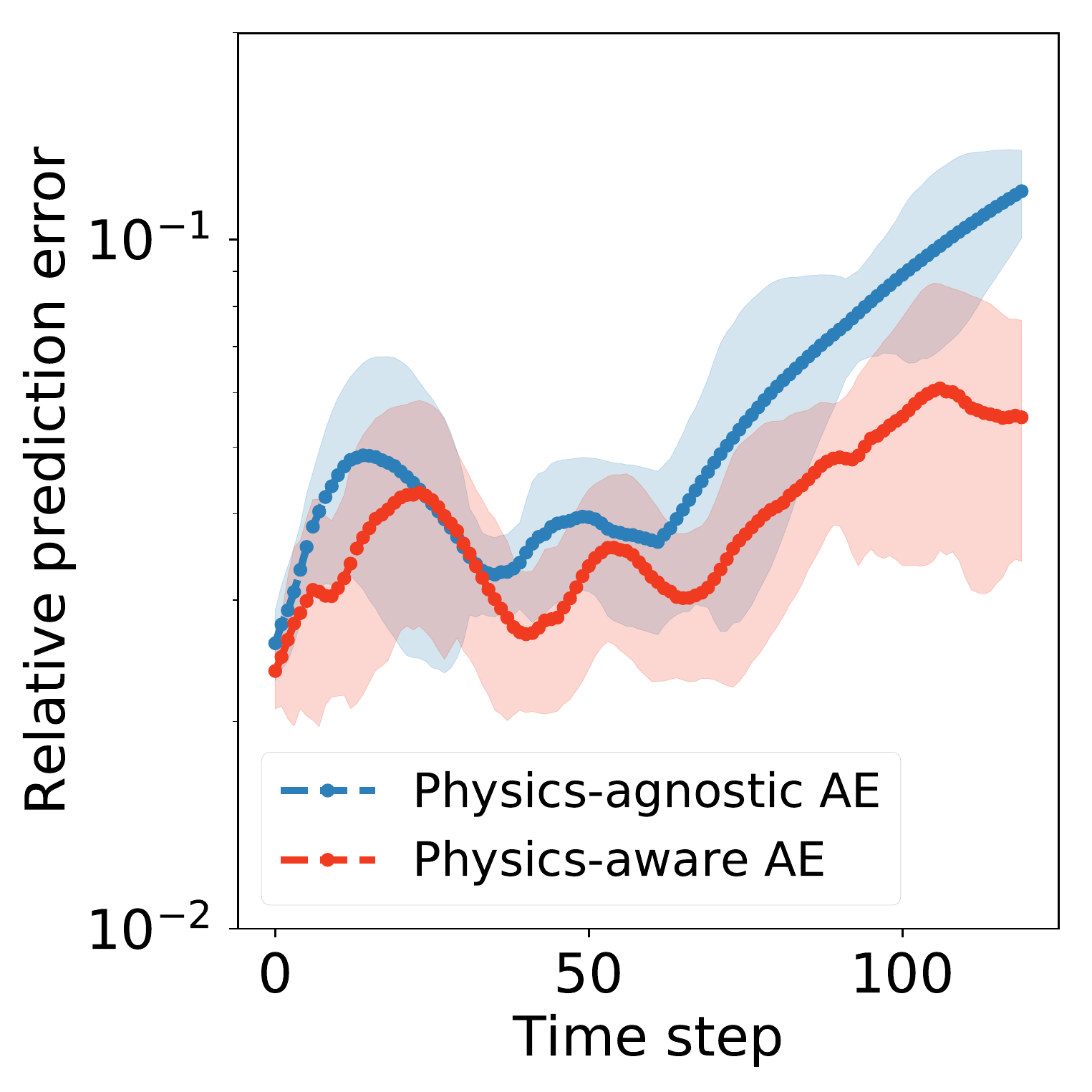}
		\caption{With LR $1\mathrm{e}{-2}$ and WD $1\mathrm{e}{-8}$.}
	\end{subfigure}
	~	
	\begin{subfigure}[t]{0.31\textwidth}
		\centering
		\DeclareGraphicsExtensions{.pdf}
		\includegraphics[width=1\textwidth]{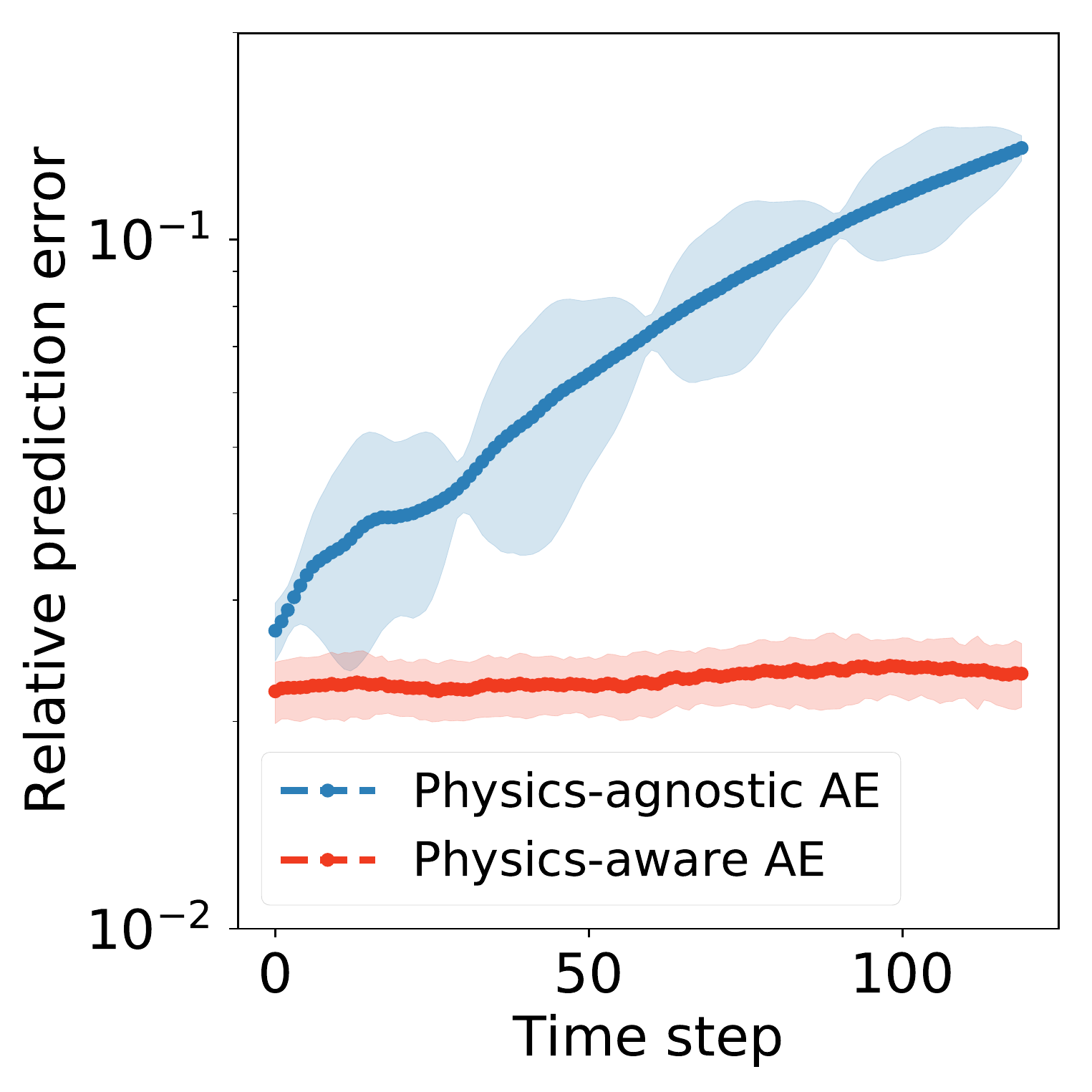}
		\caption{With LR $5\mathrm{e}{-3}$ and WD $1\mathrm{e}{-6}$.}
	\end{subfigure}	
	~	
	\begin{subfigure}[t]{0.31\textwidth}
		\centering
		\DeclareGraphicsExtensions{.pdf}
		\includegraphics[width=1\textwidth]{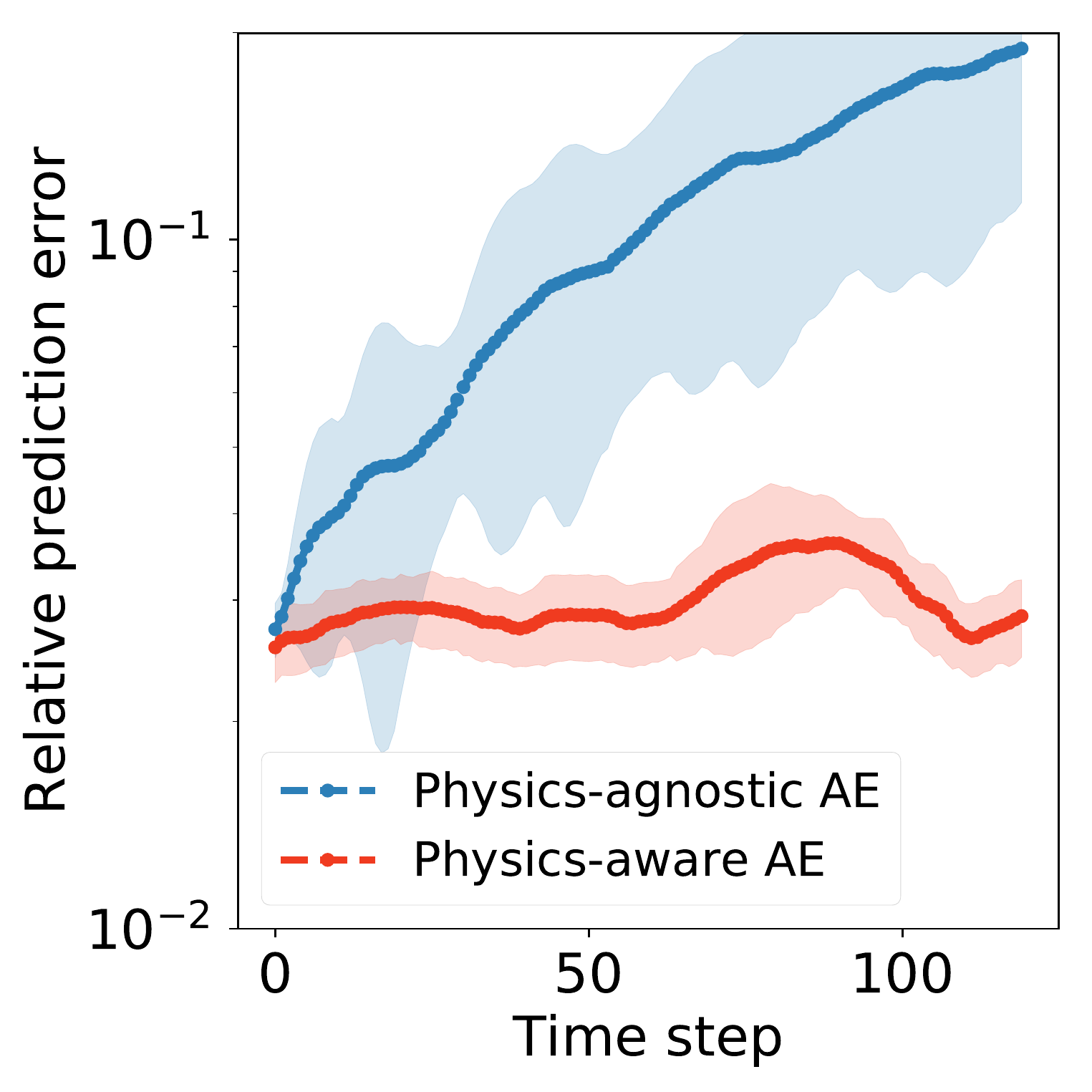}
		\caption{With LR $5\mathrm{e}{-3}$ and WD $1\mathrm{e}{-8}$.}
	\end{subfigure}	
	
	\caption{Summary of results for the flow behind a cylinder. The physics-aware model outperforms the physics-agnostic model for various tuning parameter configurations. The plots (b) and (c) show the complex eigenvalues of $\mathbf{\Omega}$, which correspond to the models in (a).  }
	\label{fig:cylinderflow_additional}
\end{figure}

\subsection{Flow behind a cylinder} 

As a canonical example, we consider a downsampled fluid flow behind a cylinder, which is characterized by a periodically shedding wake structure~\cite{Noack2003jfm}. The dataset comprises $250$ fluid flow snapshots in time, each consisting of $64\times 64$ spatial grid points. We split the sequence into a training (first $100$ snapshots) and test set (remaining $150$ snapshots).

We evaluate the quality of the physics-agnostic model (minimizing~\eqref{eq:base_loss}) and the physics-aware model (minimizing~\eqref{eq:pcl_loss}) by studying their ability to estimate future fluid flow fields.
For both models, we expect that the extrapolation in time will eventually break down, but we expect that there will be a larger range of time over which the extrapolation is valid for models that are designed to have the stability properties of the underlying physical system.
Figure~\ref{fig:cylinderflow_additional} summarizes the results for varying learning rate (LR) and weight decay (WD) configurations. 
The physics-aware model shows an improved generalization performance, when averaged over 30 initial conditions. Further, it can be seen that the stability-promoting prior reduces the prediction uncertainty.

Of course, one can also fiddle around with the amount of weight decay until all eigenvalues of $\mathbf{\Omega}$ have magnitude less than one (\ie, increasing the amount of weight decay shrinks the eigenvalues towards the origin). However, a physics-informed prior appears to be a more elegant solution as well as improves the interpretability of the model.

\begin{figure}[!b]
	\centering
	\begin{subfigure}[t]{0.31\textwidth}
		\centering
		\DeclareGraphicsExtensions{.pdf}
		\includegraphics[width=1\textwidth]{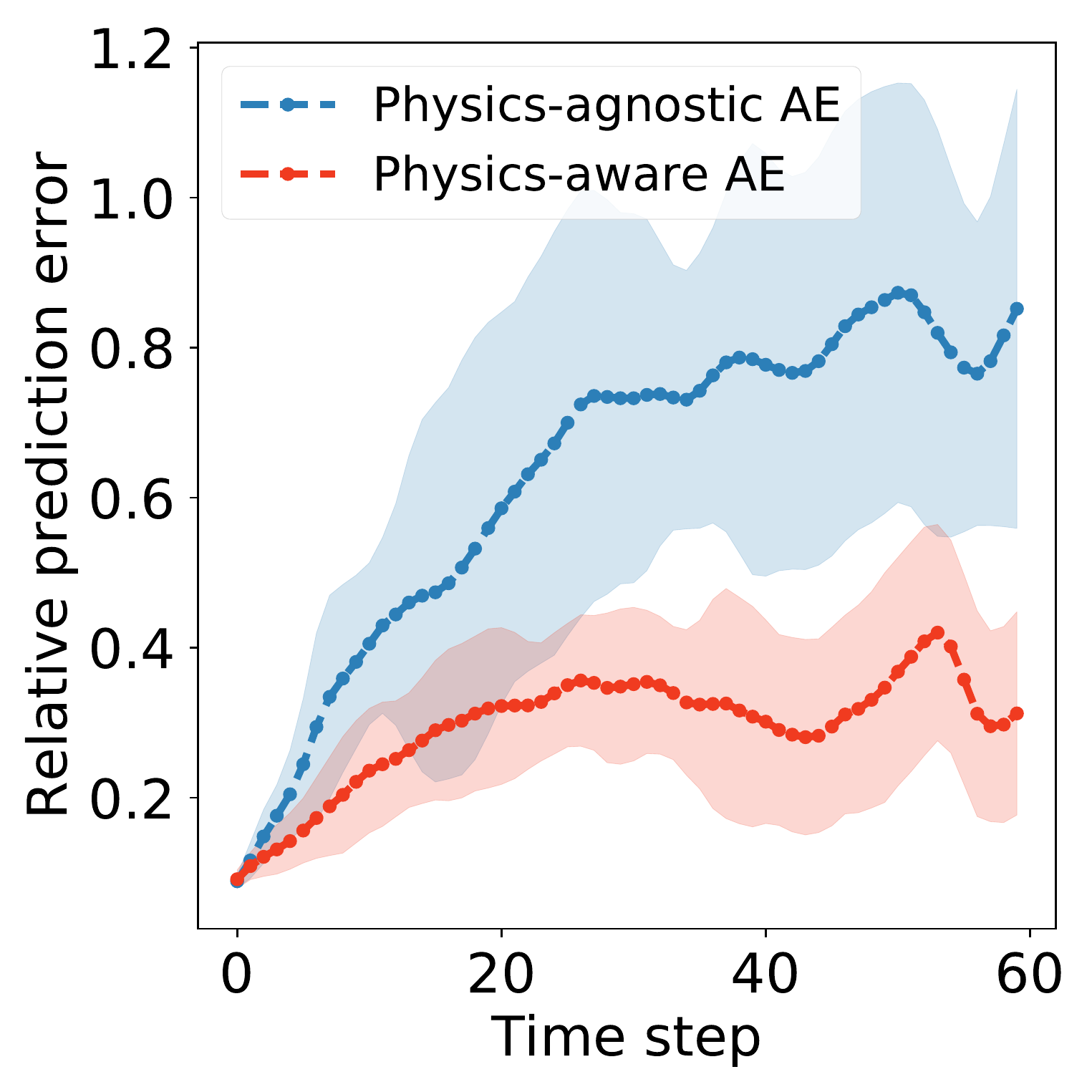}
		\caption{With LR $1\mathrm{e}{-2}$ and WD $1\mathrm{e}{-6}$.}
	\end{subfigure}
	~
	\begin{subfigure}[t]{0.31\textwidth}
		\centering
		\DeclareGraphicsExtensions{.pdf}
		\includegraphics[width=1\textwidth]{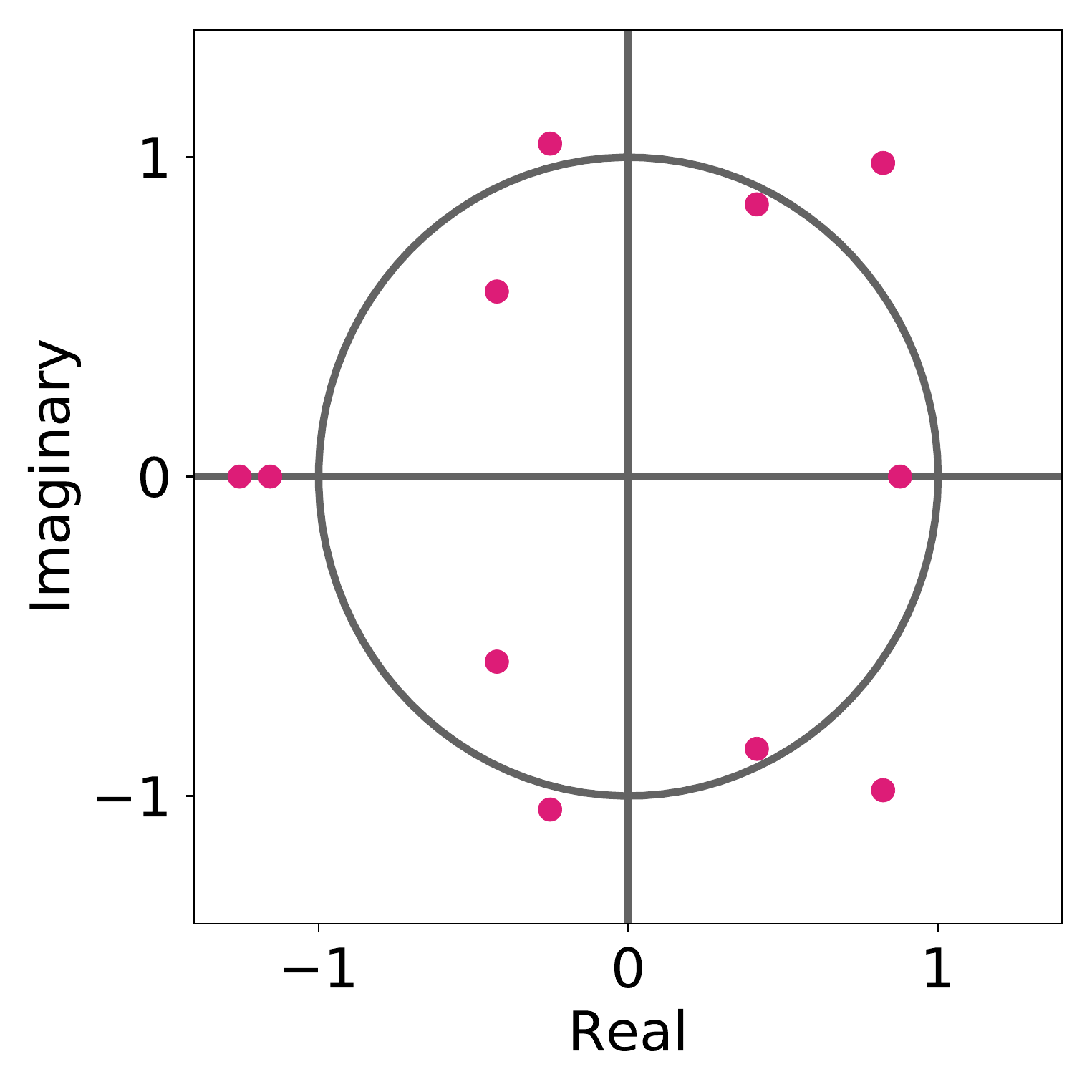}
		\caption{Physics-agnostic model.}
	\end{subfigure}\vspace{+0.1cm}
	~
	\begin{subfigure}[t]{0.31\textwidth}
		\centering
		\DeclareGraphicsExtensions{.pdf}
		\includegraphics[width=1\textwidth]{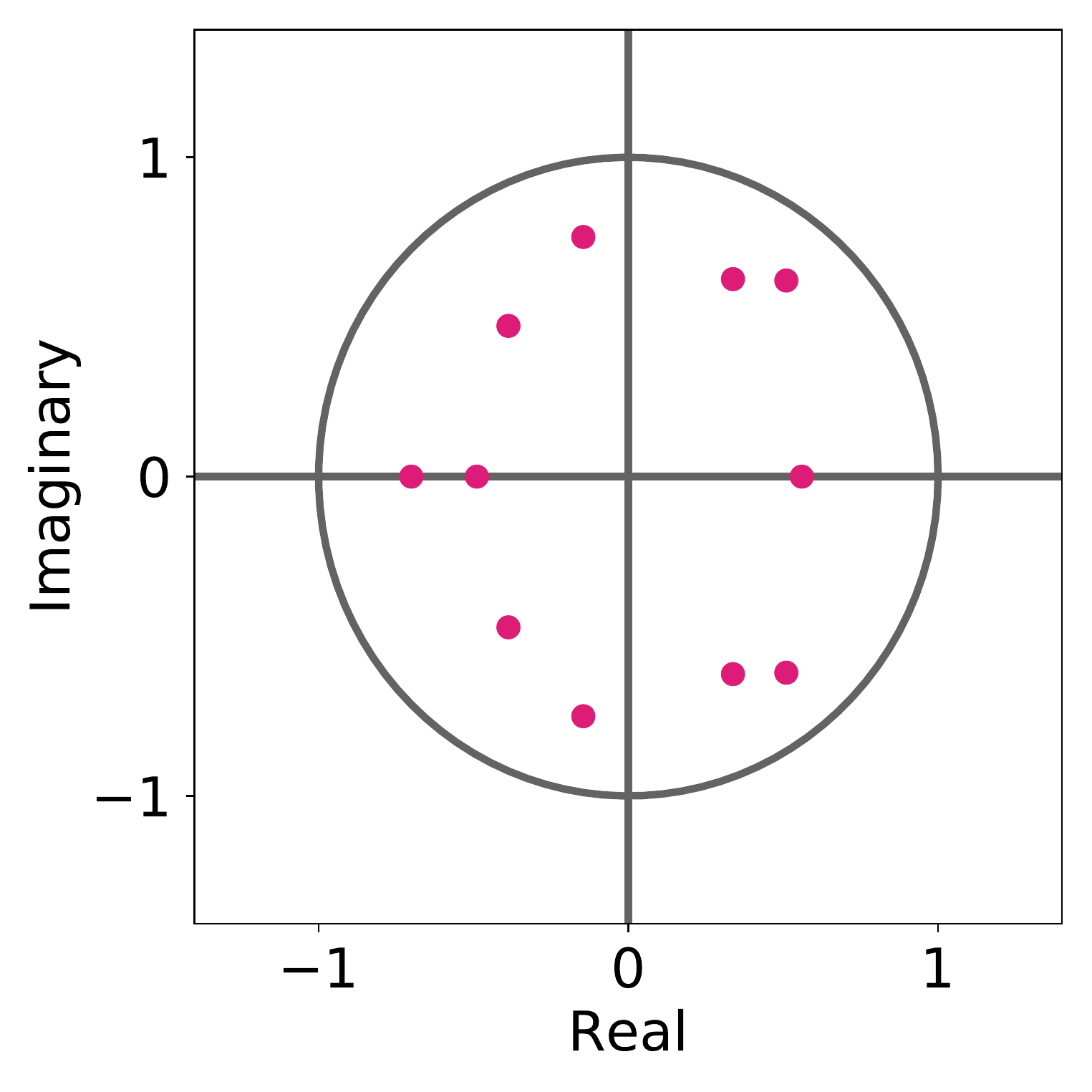}
		\caption{Physics-informed model.}
	\end{subfigure}
	
	\begin{subfigure}[t]{0.31\textwidth}
		\centering
		\DeclareGraphicsExtensions{.pdf}
		\includegraphics[width=1\textwidth]{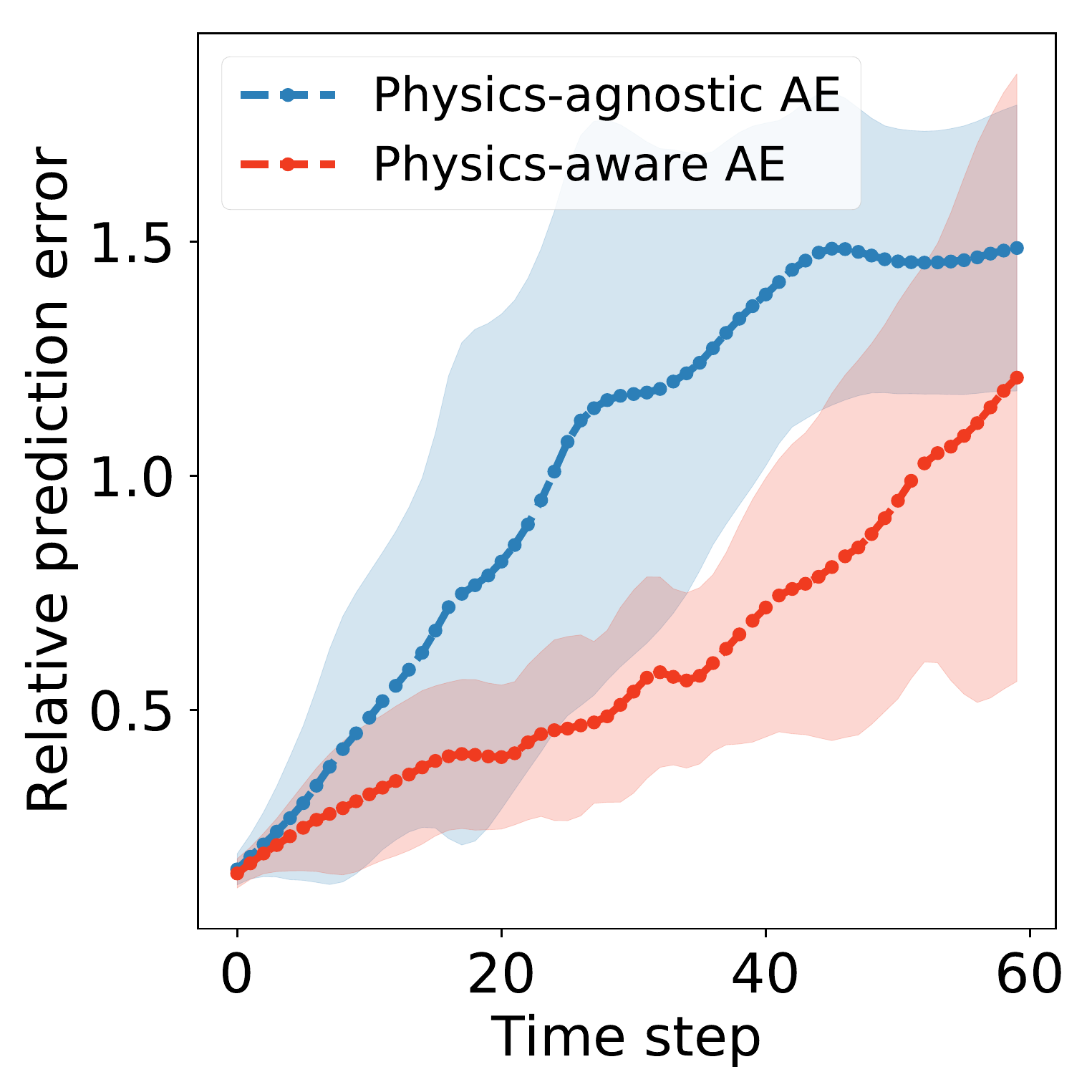}
		\caption{With LR $1\mathrm{e}{-3}$ and WD $1\mathrm{e}{-6}$.}
	\end{subfigure}
	~
	\begin{subfigure}[t]{0.31\textwidth}
		\centering
		\DeclareGraphicsExtensions{.pdf}
		\includegraphics[width=1\textwidth]{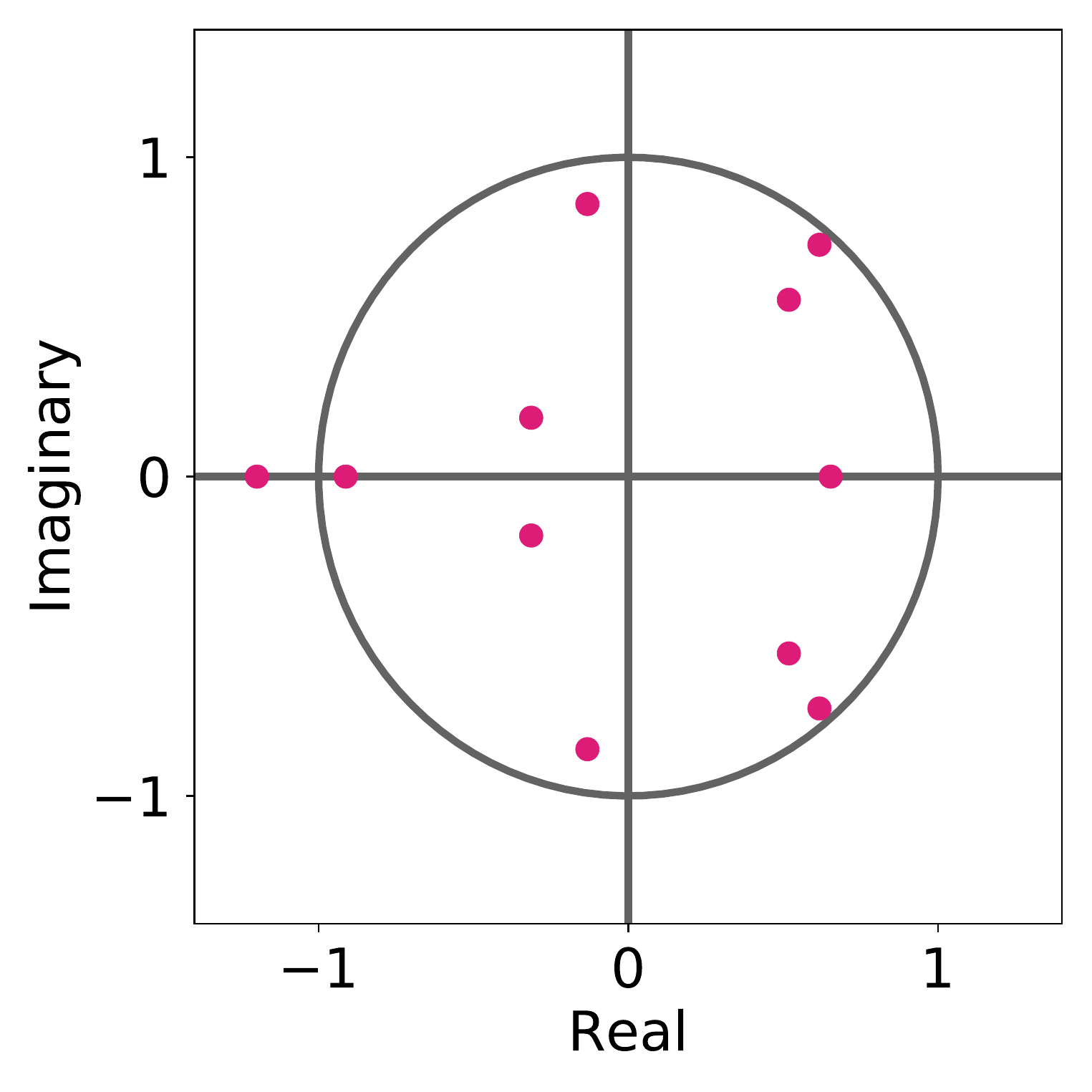}
		\caption{Physics-agnostic model.}
	\end{subfigure}
	~
	\begin{subfigure}[t]{0.31\textwidth}
		\centering
		\DeclareGraphicsExtensions{.pdf}
		\includegraphics[width=1\textwidth]{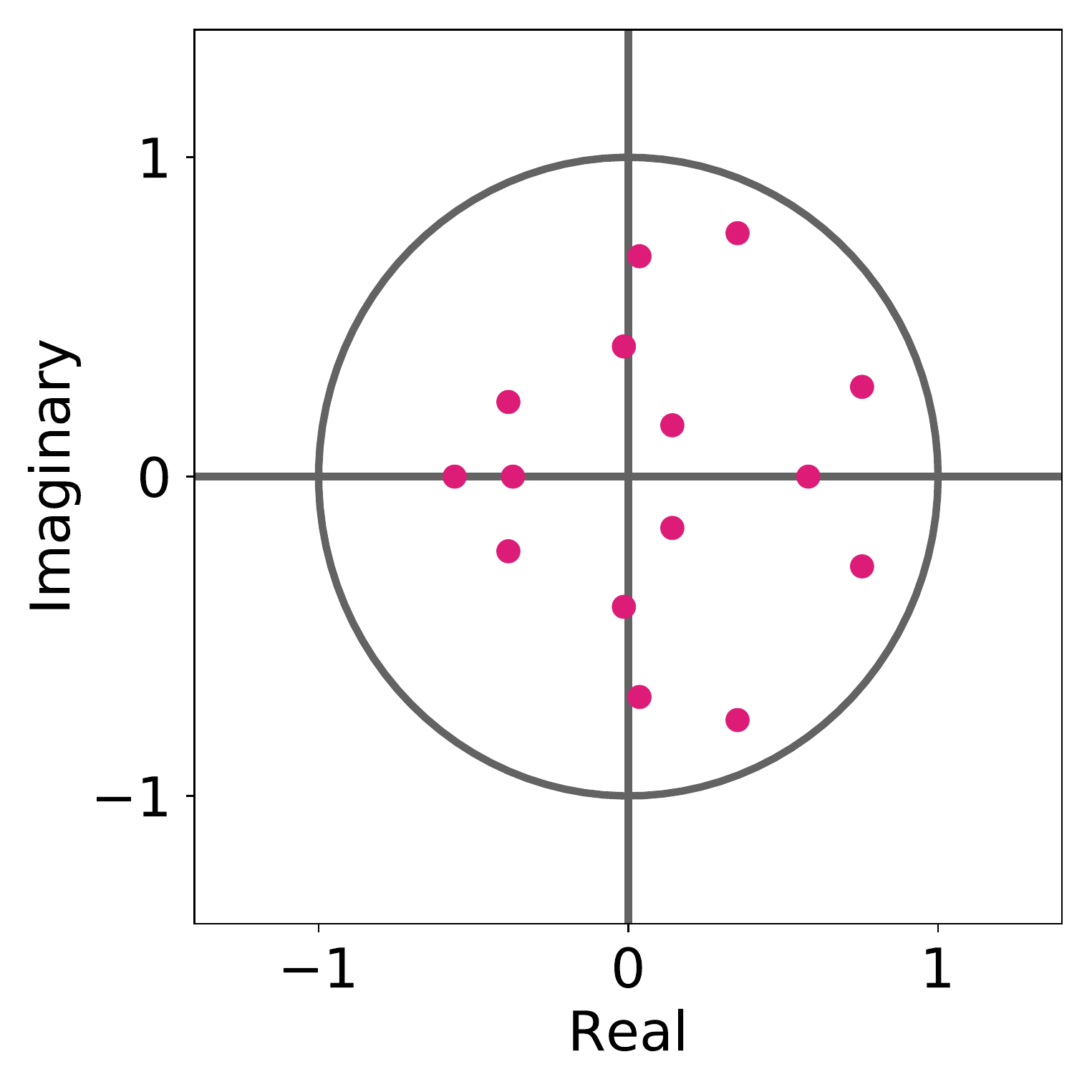}
		\caption{Physics-informed model.}
	\end{subfigure}
	
	\caption{Summary of results for the SST data. The physics-informed model shows a better generalization performance over a prediction horizon of 60 days.}
	\label{fig:sst}
\end{figure}

\subsection{Sea surface temperature of the gulf of Mexico} 

Next, we model the sea surface temperature (SST) of the the Gulf of Mexico as a real-world example to demonstrate the performance of our physics-informed autoencoder. 
The National Oceanic \& Atmospheric Administration (NOAA) at \url{http://www.esrl.noaa.gov/psd/} provides daily sea surface temperatures for the last 26 years. We consider the daily SSTs for the Gulf of Mexico over a period of six years (2012-2018). 
The data comprise $2190$ snapshots in time with spatial resolution of $64\times 64$. We split the sequence into a training (first $1825$ snapshots) and test set (remaining $365$ snapshots).

We aim to predict the fluctuations around the mean temperature. 
Figure~\ref{fig:sst} summarizes the results for two learning rate (LR) and weight decay (WD) configurations. 
Again, the physics-aware model shows an improved generalization performance for a larger range of time. 
The prediction error is overall substantially larger than in the previous examples. 
This is because predicting the fluctuations (in this non-toy model) is a challenging problem, since complex ocean dynamics lead to rich flow phenomena, featuring various seasonal fluctuations. 
It is possible that deeper neural networks would improve the prediction performance, but initial results using a residual-block-type autoencoder could not outperform the results shown here. 
Of course, for more complex models, we expect that respecting stability properties will also lead to improved performance.

\section{Conclusion}

Neural networks have been shown to be a highly valuable tool for dynamical modeling, prediction and control of fluid flows.
Surprisingly, these data-driven models have the ability to learn implicitly some of the physical properties (encoded in the data) reasonably well, if a sufficient amount of data is provided for training. 
However, often the amount of data is limited, and one has knowledge about the data generation mechanisms.
In this case, physics-informed learning might help to improve considerably the generalization performance.  
To accomplish this, we introduced a method for training autoencoders that preserve Lyapunov stability. 
This simple, yet effective, approach of including a physics-informed stability-enhancing prior into the learning process shows a substantial performance boost for several fluid flow prediction tasks.
A minor disadvantage is that we need an additional tuning parameter, but we have observed that tuning this relatively-robust parameter is not a problem.

{\normalsize
\bibliographystyle{IEEEtran} 
\bibliography{shallow} 
}

\clearpage

\appendix

\section{Network architectures}

Here, we provide details about the network architecture of the autoencoders which are use for our experiments. The networks are implemented in Python using PyTorch.
Tables~\ref{tab:arch_model1}--~\ref{tab:arch_model2} show the details. For all experiments we use a similar architecture design. The difference is that we use a slightly wider design (more neurons at the first and last layer) for the SST dataset, while using a slightly smaller dynamics layer. That is because this is a more complex problem, which benefits from the more expressive power of the wider design.

\begin{table}[!h]
	\centering\scalebox{0.9}{
		\begin{tabular}{lcccccccc} \toprule
			Type	& Layer     & Weight size   & Input Shape         & Output Shape  & Activation & Batch normalization \\
			\midrule
			Encoder	& FC   & 4,096 $\times$ 40  &   4,096    & 40   &  tanh &   - \\
			Encoder	& FC   & 40 $\times$ 25     &  40      	& 25       &  tanh & - \\
			Encoder	& FC   & 25 $\times$ 15     &  25      	& 15       &  linear & True \\
			\midrule
			Dynamics	& FC   & 15 $\times$ 15     &  15      	& 15       &  linear & - \\
			\midrule		
			Decoder	& FC   & 15 $\times$ 25     &  15      	& 25       &  tanh & - \\
			Decoder	& FC   & 25 $\times$ 40     &  25      	& 40       &  tanh & - \\			
			Decoder	& FC   & 40 $\times$ 4,096   &  40        	& 4,096   &  linear  &   - \\ \bottomrule 
	\end{tabular}}\vspace{+0.3cm}
	\caption{Architecture of the autoencoder for the flow behind a cylinder. For training we use a batch of $34$ samples.}
	\label{tab:arch_model1}
\end{table}
\begin{table}[!h]
	\centering\scalebox{0.9}{
		\begin{tabular}{lcccccccc} \toprule
			Type	& Layer     & Weight size   & Input Shape         & Output Shape  & Activation & Batch normalization \\
			\midrule
			Encoder	& FC   & 4,096 $\times$ 100  &   4,096    & 100   &  tanh &   - \\
			Encoder	& FC   & 100 $\times$ 25     &  41000      	& 25       &  tanh & - \\
			Encoder	& FC   & 25 $\times$ 15     &  25      	& 11       &  linear & True \\
			\midrule
			Dynamics	& FC   & 11 $\times$ 11     &  11      	& 11       &  linear & - \\
			\midrule		
			Decoder	& FC   & 11 $\times$ 25     &  11      	& 25       &  tanh & - \\
			Decoder	& FC   & 25 $\times$ 100     &  25      	& 100       &  tanh & - \\			
			Decoder	& FC   & 100 $\times$ 4,096   &  100        	& 4,096   &  linear  &   - \\ \bottomrule 
	\end{tabular}}\vspace{+0.3cm}
	\caption{Architecture of the autoencoder for the SST dataset. For training we use a batch of $156$ samples.}
	\label{tab:arch_model2}
\end{table}

\section{Visual results}

Here we show some visual results to support our experiments. 

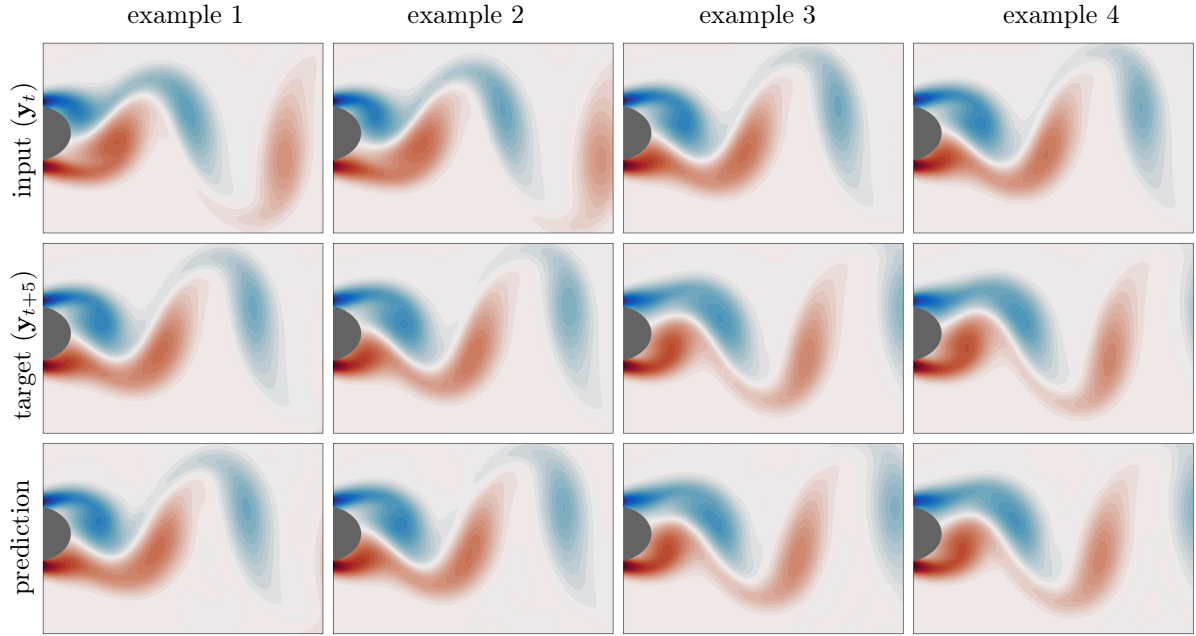
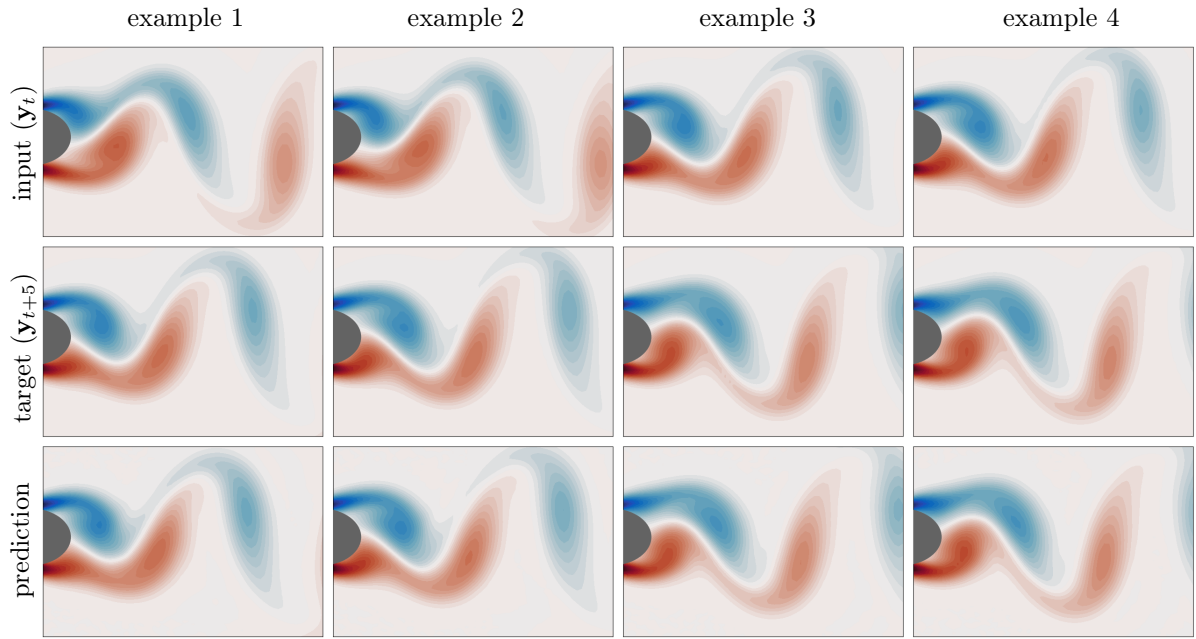
\begin{figure}[!b]
	\centering
	\DeclareGraphicsExtensions{.png}
	\vspace{+0.5cm}
	
	\begin{subfigure}[t]{0.95\textwidth}
		\centering
		\begin{overpic}[width=0.99\textwidth]{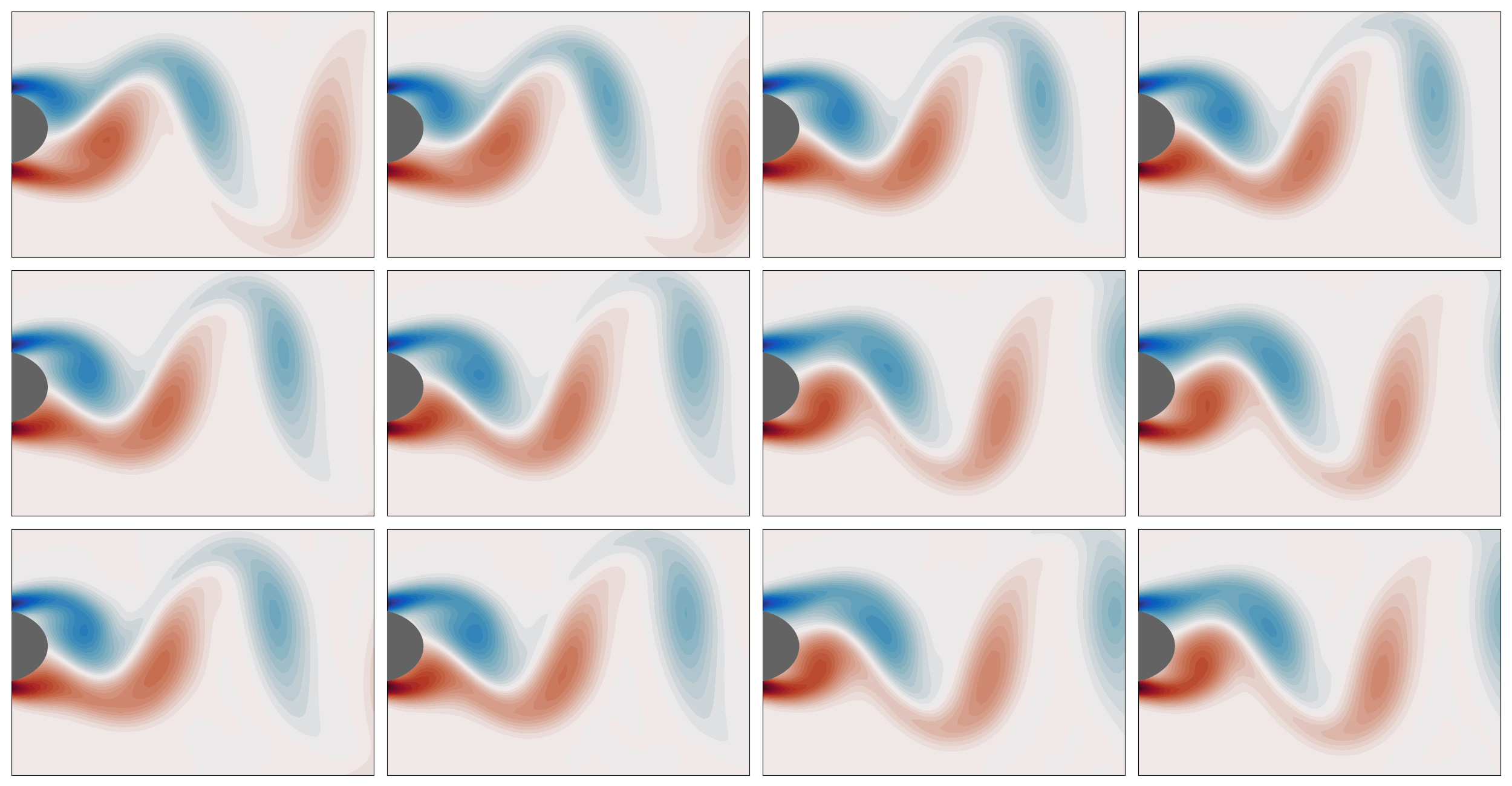} 
			\put(8,53){{\small example 1}}
			\put(32,53){{\small example 2}}
			\put(57,53){{\small example 3}}
			\put(83,53){{\small example 4}}
			\put(-2,38){\rotatebox{90}{\small input ($\mathbf{y}_t$)}}
			\put(-2,19){\rotatebox{90}{\small target ($\mathbf{y}_{t+5}$)}}
			\put(-2,4){\rotatebox{90}{\small prediction}}		
		\end{overpic}
		\caption{Physics-agnostic model.}
	\end{subfigure}

	\begin{subfigure}[t]{0.99\textwidth}
		\centering
		\vspace{+0.9cm}
		\begin{overpic}[width=0.95\textwidth]{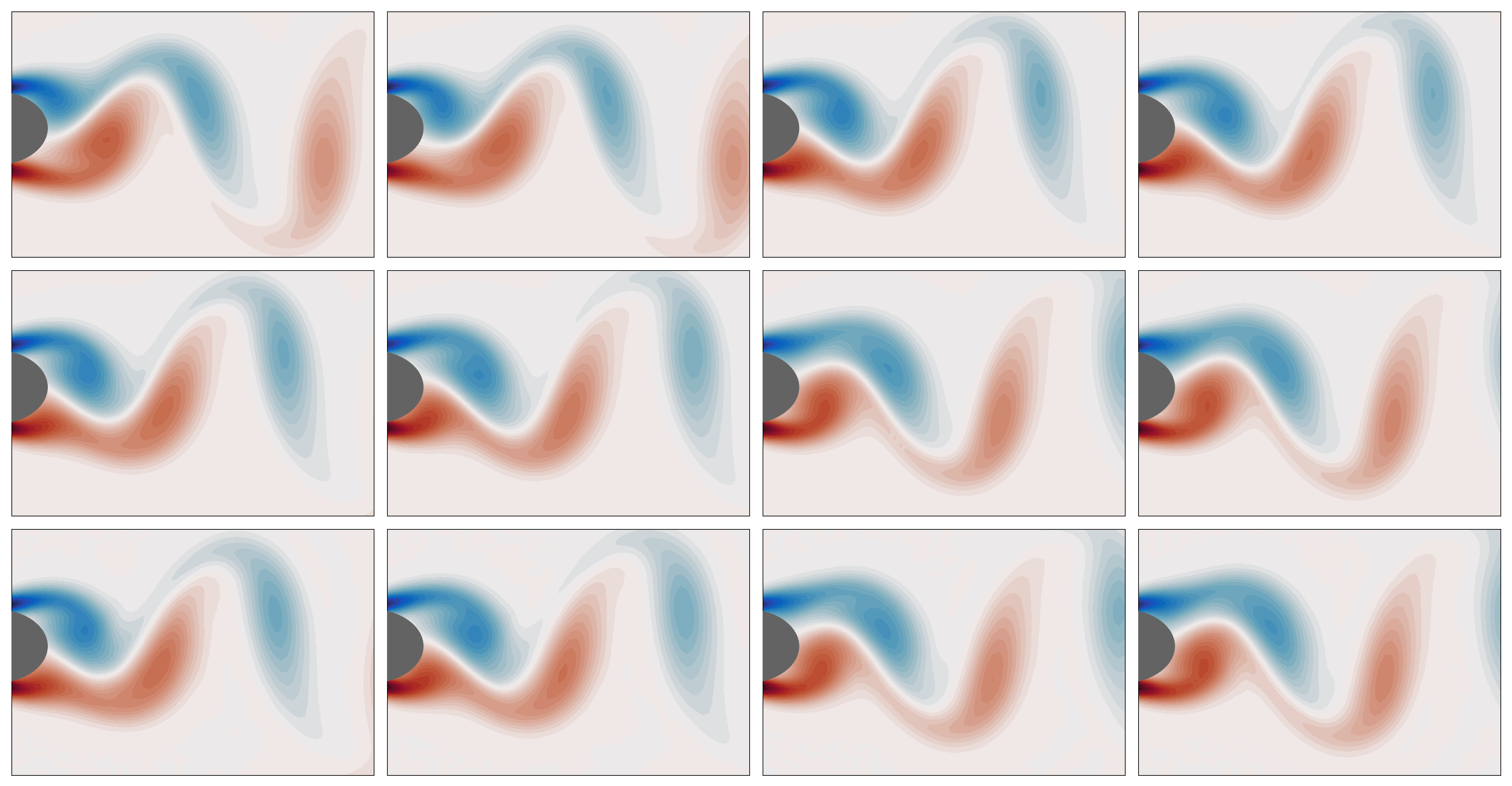} 
			\put(8,53){{\small example 1}}
			\put(32,53){{\small example 2}}
			\put(57,53){{\small example 3}}
			\put(83,53){{\small example 4}}
			\put(-2,38){\rotatebox{90}{\small input ($\mathbf{y}_t$)}}
			\put(-2,19){\rotatebox{90}{\small target ($\mathbf{y}_{t+5}$)}}
			\put(-2,4){\rotatebox{90}{\small prediction}}		
		\end{overpic}	
		\caption{Physics-aware model.}
	\end{subfigure}
	\caption{Visual results for the flow past cylinder. Predictions of future states using the physics-aware model for 4 different initial conditions. We predict the evolution of the flow field over 5 time steps.}
	\label{fig:flow_visual_results}
\end{figure}

Figure~\ref{fig:flow_visual_results} shows examples for the flow behind a cylinder. The top rows show four different initial conditions and the middle rows a future target snapshot (i.e., the evolved flow field over 15 time steps). The predictions of our physics-aware model are shown in Figure~\ref{fig:flow_visual_results} (b). Here we consider the models trained with learning rate and weight decay set to $1e-2$ and $1e-6$, respectively. By visual inspection we can see that the estimated snapshot closely matches the true target. In this simple case both the physics-aware and agnostic model are able to achieve a good generalization performance over a short prediction horizon. We will see below that the prediction performance is distinct when we consider more complex problems.  

\begin{figure}[!t]
	\centering
	\begin{subfigure}[t]{0.40\textwidth}
		\centering
		\DeclareGraphicsExtensions{.png}
		\includegraphics[width=1\textwidth]{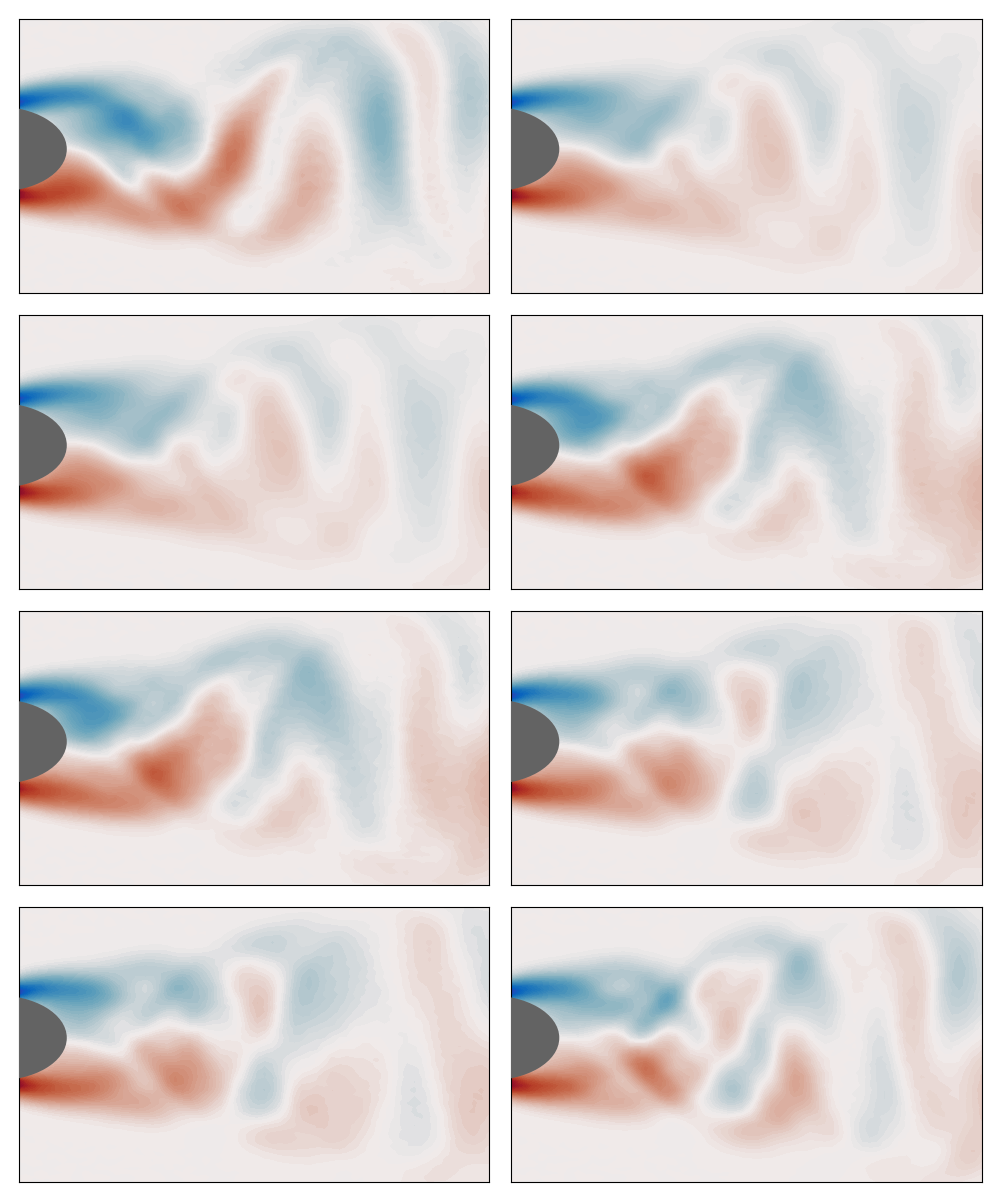}
		\caption{Physics-agnostic model.}
	\end{subfigure}
	~
	\begin{subfigure}[t]{0.40\textwidth}
		\centering
		\DeclareGraphicsExtensions{.png}
		\includegraphics[width=1\textwidth]{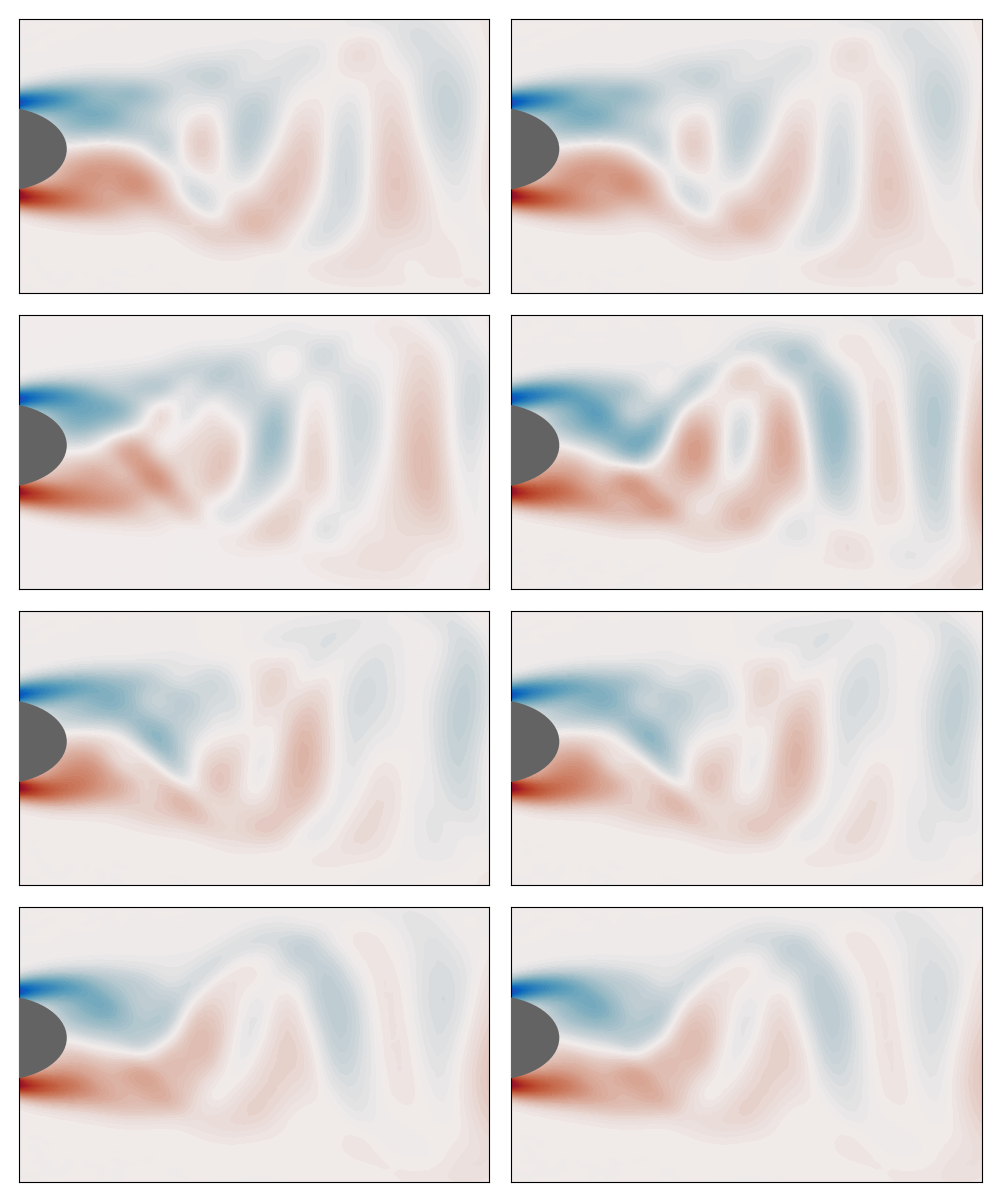}
		\caption{Physics-aware model.}
	\end{subfigure}
	
	\caption{Reconstructed eigenvectors of the dynamics $\mathbf{\Omega}$ for both the physics-agnostic and the physics-aware model. }
	\label{fig:flow_vector}
\end{figure}

Next, Figure~\ref{fig:flow_vector} shows the reconstructed eigenvectors of the dynamics $\mathbf{\Omega}$ for both the physics-agnostic and aware model. It can be seen that the two models learn to encoder different dynamics. We can see that the eigenvectors of the physics-aware model have more structure than those of the physics-agnostic model. 

\begin{figure}[!t]
	\centering
	\vspace{+1.5cm}
	\begin{subfigure}[t]{0.48\textwidth}
		\centering
		\DeclareGraphicsExtensions{.png}
		\begin{overpic}[width=0.98\textwidth]{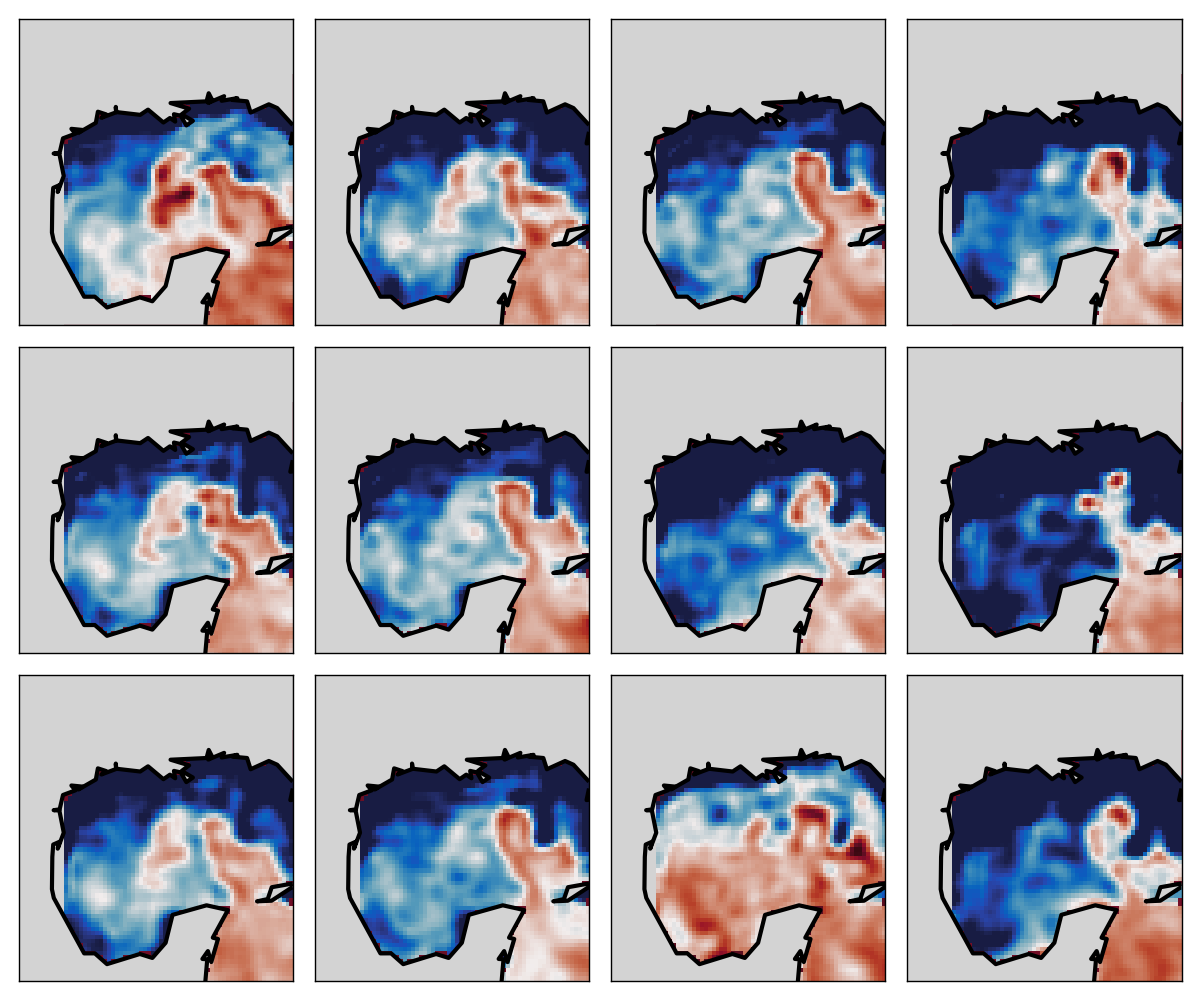} 
			\put(4.5,85){{\small example 1}}
			\put(28,85){{\small example 2}}
			\put(53,85){{\small example 3}}
			\put(77,85){{\small example 4}}
			\put(-4,60){\rotatebox{90}{\small input ($\mathbf{y}_t$)}}
			\put(-4,30){\rotatebox{90}{\small target ($\mathbf{y}_{t+5}$)}}
			\put(-4,4){\rotatebox{90}{\small prediction}}		
		\end{overpic}		
		
		\caption{Physics-agnostic model.}
	\end{subfigure}
	~
	\begin{subfigure}[t]{0.48\textwidth}
		\centering
		\DeclareGraphicsExtensions{.png}
		\begin{overpic}[width=0.98\textwidth]{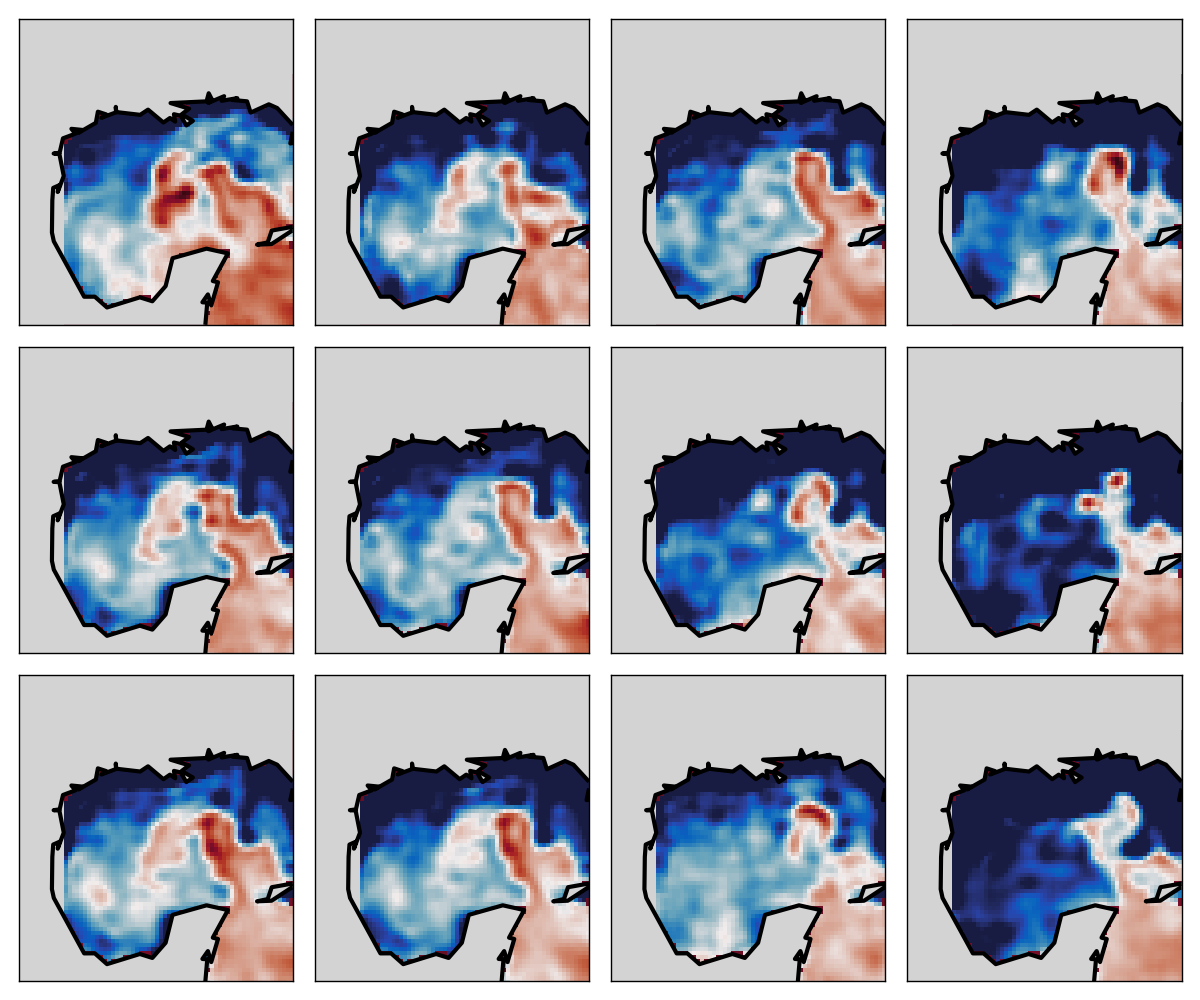} 
			\put(4.5,85){{\small example 1}}
			\put(28,85){{\small example 2}}
			\put(53,85){{\small example 3}}
			\put(77,85){{\small example 4}}
			\put(-4,60){\rotatebox{90}{\small input ($\mathbf{y}_t$)}}
			\put(-4,30){\rotatebox{90}{\small target ($\mathbf{y}_{t+5}$)}}
			\put(-4,4){\rotatebox{90}{\small prediction}}		
		\end{overpic}	
		\caption{Physics-aware model.}
	\end{subfigure}
	
	\caption{Visual results for the SST data. Predictions of future states using both the physics-agnostic and physics-aware model for 4 different initial conditions. We predict the evolution of the field over 5 time steps. }
	\label{fig:visual_sst}
\end{figure}
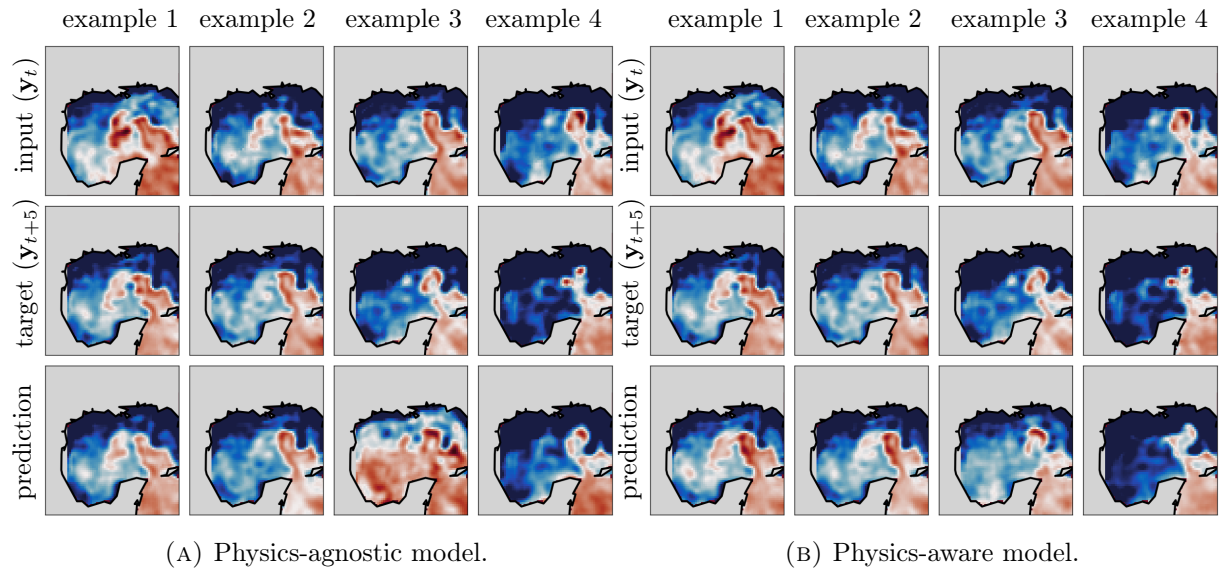

Figure~\ref{fig:visual_sst} shows visual results for the prediction performance for the climate data. Here, we extrapolate the fluctuations around the mean temperature for 5 time steps into the future. It can be see that the physics-aware model shows a better approximation quality for the extrapolated temperature fields than the physics-agnostic model. In particular, the difference is distinct for example 3.

\end{document}